\documentclass[a4paper,11pt]{article}

\usepackage{jheppub} 
\usepackage{lineno}

\usepackage{multirow}
\usepackage{placeins}
\usepackage{diagbox}

\newcommand{\beaa}{\begin{eqnarray*}}
	\newcommand{\enaa}{\end{eqnarray*}}
\newcommand{\bea}{\begin{eqnarray}}
	\newcommand{\ena}{\end{eqnarray}}
\newcommand{\seq}{\begin{subequations}}
	\newcommand{\sen}{\end{subequations}}
\newcommand{\eq}{\begin{eqnarray}}
	\newcommand{\en}{\end{eqnarray}}

\def\shiftdown#1{#1\llap{\lower.04ex\hbox{#1}}}

\def\nn{\nonumber}

\def \be  {\begin{equation}}
	\def \ee  {\end{equation}}
\def \ba  {\begin{eqnarray}}
	\def \ea  {\end{eqnarray}}
\def \baa {\begin{eqnarray*}}
	\def \eaa {\end{eqnarray*}}

\def \bb  {\begin{thebibliography}}
	\def \eb  {\end{thebibliography}}

\def \lab #1 {\label{#1}}

\def \e {\mbox{e}}

\title{Angular coefficients of the Drell-Yan process across different rapidity and kinematical ranges}

\author[a,b]{Valery E. Lyubovitskij,} 
\author[c,d]{Alexey S. Zhevlakov,}
\author[c,e]{Iurii A. Anikin}

\affiliation[a]{Institut f\"ur Theoretische Physik, Universit\"at T\"ubingen,\\Auf der Morgenstelle 14, D-72076 T\"ubingen, Germany}
\affiliation[b]{Millennium Institute for Subatomic Physics at the High-Energy Frontier (SAPHIR) of ANID,\\Fern\'andez Concha 700, Santiago, Chile}  
 
\affiliation[c]{Bogoliubov Laboratory of Theoretical Physics, Joint Institute for Nuclear Research,\\ 141980 Dubna, Russia} 

\affiliation[d]{Matrosov Institute for System Dynamics and 
Control Theory SB RAS,\\Lermontov str., 134, 664033, Irkutsk, Russia}

\affiliation[e]{Department of Physics, Tomsk State University,\\634050 Tomsk, Russia}

\emailAdd{valeri.lyubovitskij@uni-tuebingen.de}
\emailAdd{zhevlakov@theor.jinr.ru}
\emailAdd{iuanikin@theor.jinr.ru}

\abstract{
We present comprehensive analysis of the angular structure of 
the Drell-Yan process at different rapidity and kinematical ranges 
using data of the ATLAS, LHCb, and CMS Collaborations at CERN LHC. 
From theory side we discuss next-to-leading order calculations  
in the framework of the collinear perturbative QCD and geometrical method.
}


\begin{document}

\maketitle

\flushbottom


\section{Introduction}

Discovery of composite structure of hadrons 
at SLAC in 1960-th supported by theoretical progress (quark model building, Feynman parton model, etc.) stimulated comprehensive study of hadron physics. Drell-Yan (DY) processes is one of the powerful and key methods for study of perturbative and nonperturbative structure of hadrons. Study of DY processes gives  a unique opportunity to shed light on strong interaction properties -- 
perturbative (resummation of soft and collinear singularities) and nonperturbative (extraction of parton distributing functions (PDFs) of quarks 
and gluons in composite hadrons). 
Different types of PDFs being one of fundamental properties of hadron constituents describe  distribution of quarks and gluons in hadrons (nuclei) and encode information about longitudinal hadron momentum, transverse momenta of partons, and spin correlations of hadrons and partons. 
Theoretical study of the DY processes is motivated by currently 
running and planned precise experiments at world-wide facilities~(see, e.g., 
refs.~\cite{CMS:2015cyj,ATLAS:2016rnf,LHCb:2022tbc,COMPASS:2017jbv,COMPASS:2023vqt,Adams:2018pwt,NuSea:2006gvb,CDF:2011ksg,STAR:2015vmv,PHENIX:2018dwt,SeaQuest:2019hsx,Accardi:2012qut}). 

The LHC experiments (CMS~\cite{CMS:2015cyj}, ATLAS~\cite{ATLAS:2016rnf}, and LHCb~\cite{LHCb:2022tbc}) can be 
considered as a main provider of DY data using $pp$ collisions with 
possibility of study different ranges of invariant mass $Q$, transverse momentum $Q_T$, and rapidity of the lepton pair $y$. 
Analysis of lepton angular structure of DY processes (in particular, angular coefficients) gives direct access to information about inner structure of hadrons. The knowledge of angular coefficients also plays 
an important role in more precise determination of such parameters 
as the $W$ boson mass $m_W$ or Weinberg angle $\theta_W$.  
For recent overview of the current status of
quantum chromodynamics (QCD) precision predictions for DY processes, 
see, e.g., ref.~\cite{Alekhin:2024mrq}. 

Latest advanced calculations of the DY angular distributions and coefficients provide a crucial opportunity for high-precision tests of 
the electroweak sector of the Standard Model (SM). 
Analytical studies of angular coefficients were done in 
next-to-leading (NLO) in ref.~\cite{Boer:2006eq,Berger:2007jw,Lyubovitskij:2024jlb} 
and in next-to-next-to-leading (NNLO)~\cite{Mirkes:1992hu}.  
A comprehensive comparison of the available experimental data for 
the Drell-Yan lepton angular coefficients $\lambda$ and $\nu$ to calculations at 
NLO and NNLO of perturbative QCD (pQCD) has been presented in ref.~\cite{Lambertsen:2016wgj}. 
Notably, the precision of both inclusive and fully differential DY cross sections has been extended  NNLO~\cite{Hamberg:1990np,vanNeerven:1991gh,%
Anastasiou:2003yy,Melnikov:2006di,Catani:2009sm,Gavin:2010az,Gavin:2012sy} 
to N$^3$LO~\cite{Duhr:2020seh,Baglio:2022wzu,Camarda:2021jsw}. 
A careful and consistent inclusion of electroweak corrections has been carried out in refs.~\cite{Baur:2001ze,%
Dittmaier:2001ay,Arbuzov:2005dd,Arbuzov:2007db,CarloniCalame:2007cd}.  
At present time we need to perform a precise and fast theoretical calculations, 
which are provided by different existing Monte Carlo (MC) generators. 
In particular, DY processes were analyzed using NNLOjet~\cite{Gauld:2017tww}, Pythia~\cite{Martin:2004dh}, Herwig \cite{Bellm:2015jjp}, DYNNLO~\cite{Catani:2009sm}, FEWZ~\cite{Gavin:2010az,Li:2012wna}. In general, the coefficients are neither extracted directly at calculation in the MC event generators. However, they can be extracted from the angular distribution shapes using the method proposed in ref.~\cite{Mirkes:1994eb}, due to the orthogonality of the polynomials. Using analytical 
results~\cite{Berger:2007jw,Lyubovitskij:2024jlb} we can obtain angular coefficients behavior without using methods based on orthogonality of the polynomials. 

Utilizing NNLOjet MC generator~\cite{Gauld:2017tww}  
the DY angular coefficients $A_{0,1,2,3,4}$ were calculated, compared with 
data provided by the CMS and ATLAS Collaborations and made predictions 
for the LHCb experiment. These studies include N$^3$LO computation of 
the angular coefficients using the method proposed in ref.~\cite{Mirkes:1994eb}. 
Besides, data were managed by Pythia~\cite{Martin:2004dh}, Herwig~\cite{Bellm:2015jjp}, DYNNLO~\cite{Catani:2009sm}, FEWZ~\cite{Gavin:2010az,Li:2012wna} in the 
experimental papers~\cite{CMS:2015cyj,ATLAS:2016rnf,LHCb:2022tbc}. 
These analysis showed an importance of the NNLO and higher-order 
corrections to the DY angular distributions/coefficients. In particular, 
it is crucial for the $A_1$ and $A_2$ coefficients that provide an adjustment 
for sensitive agreement with data and violation of the Lam-Tung (LT) 
relation~\cite{Lam:1978pu,Gauld:2017tww}. Besides, in ref.~\cite{Gauld:2017tww} 
it was noted that the N$^3$LO $O(\alpha_s^3)$ corrections are observed to have 
an important influence on the predicted shapes of the $A_0$, $A_1$, 
and $A_2$ coefficients distributions. However, it is small for the $A_3$ and $A_4$ distributions and not sensitive for prediction at existed experimental level of 
accuracy. Apart from it, one needs to note that the DY angular coefficients is a good 
tool for extraction of PDFs~\cite{Accomando:2019vqt}, 
in particular, one that done by using the open-source fit 
platform \texttt{xFitter}~\cite{Alekhin:2014irh} for Higgs production~\cite{Amoroso:2020fjw}.

Other possibilities for predictions of angular coefficients were proposed in 
refs.~\cite{Argyres:1982kg,Chang:2017kuv,Lyu:2020nul},  where authors were based on geometrical method. This   method is based on the idea to use parametrization for $Q_T$ dependence of the angular coefficients induced by the $q\bar{q}$ and $qg$ subprocesses 
at NLO. In this way, the set of free parameters is fixed using data. 
Such scheme provides a simple way for prediction 
of the $Q_T$ behavior of the angular coefficients. 
Looking at this method two main questions arise: 
(1) How accurately does this method capture the behavior of the angular coefficients for DY process; 
(2) Can we determine angular coefficients from 
rotation of the hadronic plane relative to the quark plane using $q\bar{q}\to l\bar{l}$ cross section. 
It will be tested using calculation in pQCD in 
the framework of collinear approximation~\cite{Lyubovitskij:2024jlb}  and analyzing existing data from LHC experiments. This calculation include also only NLO contribution because it is needed for comparison with geometrical method. In future we plan to expand our calculations on NNLO. 

The paper is organized as follows. 
In section~\ref{Sec2_Theory} we give a brief overview 
of the DY process and its basics blocks -- angular coefficients. 
Later we also briefly discuss dependencies of angular coefficients in the geometrical method and results in framework of collinear approximation pQCD in NLO calculation. In section~\ref{Sec3_comparison} we present statistical analysis of both approaches (our and geometrical) in comparison with available data from the LHC experiments  for angular coefficients. In section~\ref{Sec4_Summary} we summarize our results. 

\section{Theory}
\label{Sec2_Theory}

\subsection{Basics of the Drell-Yan reactions}
\label{subsec_basics of DY}

We consider the process of lepton pair production in 
the inelastic DY reaction $p(p_1)+p(p_2) \to \gamma^*/Z +X \to l_1(k_1)+l_2(k_2)+X$ with intermediate photon $\gamma^*$ and neutral weak $Z$ boson. 
The cross section of this reaction can be presented as the contraction of leptonic $L_{\mu\nu}$ and hadronic $W_{\mu\nu}$ tensors: 
\eq 
\frac{d\sigma}{d\Omega \, d^4q} = \frac{\alpha^2}{2 (2\pi)^4 Q^4 s^2} \, L_{\mu\nu} \, W^{\mu\nu}\,,
\en 
where 
$s=(P_1 + P_2)^2$ is the hadron-level total energy, 
$\alpha = 1/137.036$ is the electromagnetic 
fine-structure constant, and $Q^2 = q^2$ is 
the time-like vector boson momentum squared. 
Hadronic momenta $P_1$ and $P_2$ are chosen in the CS frame
and related to the parton momenta 
$p_i = \xi_i P_i$, where $\xi_i$ is the partonic momentum
fraction,  $Q^+$, $Q^-$, $Q_T$ are the gauge boson
longitudinal and transverse momentum components, respectively,
with $Q^\pm = x_{1,2} \sqrt{s/2} = e^{\pm y} \sqrt{(Q^2+Q_T^2)/2}$. 
We introduce the following notations:
$\rho = Q_T/Q$ is the ratio of the transverse component and magnitude
$Q=\sqrt{Q^2}$ of the vector boson momentum, 
$x_{1,2} = 2 P_{2,1}q/s$ are the momentum fractions of the light-cone
components of the finale vector boson, 
$y = (1/2) \log(x_1/x_2)$ is the rapidity. 
We also define the $x^0_i$ fraction factors at $Q_T^2 = 0$ as
$x_{1,2}^0 = e^{\pm y} \, Q/\sqrt{s}$.  

The DY lepton angular distribution $dN/d\Omega$  
depends on the polar $\theta$ and azimuthal $\phi$ 
angles and can be conventionally expanded 
in terms of the nine helicity structure functions, 
corresponding to the specific polarization of gauge boson.  
These structure functions play important role 
and encode the polar and azimuthal asymmetries 
occurring in the DY process. 
In the CS frame the DY lepton angular 
distribution reads~\cite{Lam:1978pu,Collins:1977iv,Mirkes:1992hu,%
Boer:2006eq,Berger:2007jw,Lyubovitskij:2024civ, Lyubovitskij:2024jlb} 
\eq\label{dNdOmega} 
\frac{dN}{d\Omega} &=& \frac{d\sigma}{d\Omega d^4q} \,
\biggl(\frac{d\sigma}{d^4q}\biggr)^{-1} \!\!\!
= \frac{3}{8 \pi (2 W_T + W_L)} \,
\biggl[ 
      g_T             \, W_T
\,+\, g_L             \, W_L 
\,+\, g_\Delta         \, W_\Delta
\,+\, g_{\Delta\Delta}  \, W_{\Delta\Delta}
\nonumber\\
&+&   g_{T_P}          \, W_{T_P}
\,+\, g_{\nabla_P}      \, W_{\nabla_P}
\,+\, g_{\nabla}        \, W_{\nabla}
\,+\, g_{\Delta\Delta_P} \, W_{\Delta\Delta_P} \,
\,+\, g_{\Delta_P}      \, W_{\Delta_P} \biggr] \,,
\en
where
$g_i = g_i(\theta,\phi)$ are the angular coefficients 
\begin{align}
\label{gcoeffs}
g_{_T} &= 1 + \cos^2\theta\,,  &
g_{_L} &= 1 - \cos^2\theta\,,  &
g_{_{T_P}} &= \cos\theta\,, \nonumber\\
g_{_{\Delta\Delta}} &= \sin^2\theta \, \cos 2\phi\,, &
g_{_{\Delta}} &= \sin 2\theta \, \cos\phi\,, &
g_{_{\nabla_P}} &= \sin\theta \, \cos\phi\,,\nonumber \\ 
g_{_{\Delta\Delta_P}} &= \sin^2\theta \, \sin 2\phi\,, &
g_{_{\Delta_P}} &= \sin 2\theta \, \sin\phi\,, &
g_{_{\nabla}} &= \,\sin\theta \, \sin\phi \,.
\end{align}

There is equivalent parametrization of the $dN/d\Omega$  
using the set of the angular coefficients $A_i$ 
in literature~\cite{Lam:1978pu,Collins:1977iv,Mirkes:1992hu,%
Boer:2006eq,Berger:2007jw}
\eq\label{eq4-sf}
\frac{dN}{d\Omega} &=& \frac{3}{16\pi} \, 
\biggl( 
1 + \cos^2\theta 
+ \frac{A_0}{2} (1-3\cos^2\theta) 
+ A_1 \sin 2\theta  \cos\phi 
+ \frac{A_2}{2} \sin^2\theta  \cos 2\phi \quad\quad\quad\quad\nonumber\\ 
&+& A_3 \sin\theta  \cos\phi 
+ A_4 \cos\theta  
+ A_5 \sin^2\theta \sin 2\phi 
+ A_6 \sin 2\theta \sin\phi 
+ A_7 \sin\theta  \sin\phi  
\biggr)\,,
\en
where 
\begin{align}
\label{A_Wrelations} 
A_0 &= \frac{2W_L}{2W_T+W_L} \,, & 
A_1 &= \frac{2W_\Delta}{2W_T+W_L} \,, &
A_2 &= \, \frac{4W_{\Delta\Delta}}{2W_T+W_L} \,, 
\nonumber\\[2mm] 
A_3 &= \frac{2W_{\nabla_P}}{2W_T+W_L} \,, &
A_4 &= \frac{2W_{T_P}}{2W_T+W_L} \,, &
A_5 &= \frac{2W_{\Delta\Delta_P}}{2W_T+W_L}  \,, 
\nonumber\\[2mm] 
A_6 &= \frac{2W_{\Delta_P}}{2W_T+W_L} \,, &
A_7 &= \, \frac{2W_{\nabla}}{2W_T+W_L} \,. 
\end{align} 
We also define  the forward-backward (FB) asymmetry coefficient $A_{\rm FB}$ 
and the convexity (transverse-longitudinal hadronic structure) asymmetry  
$A_{\rm conv}$~\cite{Lyubovitskij:2024jlb}:    
\eq
A_{\rm FB} = \frac{3}{8} \, A_4 \,, \quad 
A_{\rm conv} = \frac{3}{8} \, (2 - 3 A_0) \,. 
\label{Afb_Acon}
\en

Hadronic structure functions $W_j=W_j(x_1,x_2)$ 
characterizing DY process with colliding
hadrons $H_1(p_1)$ and $H_2(p_2)$ 
are related to the partonic-level structure functions
$w_j^{ab}(x_1,x_2)$ by the QCD collinear 
factorization formula~\cite{Lyubovitskij:2024civ} 
\eq\label{factorization}
W_j(x_1,x_2) =
\frac{1}{x_1 x_2} \, \sum\limits_{a,b}
\, \int\limits_{x_1}^1 \, dz_1
\, \int\limits_{x_2}^1 \, dz_2
\ w_j^{ab}(z_1,z_2) 
\, f_{a/H_1}\Big(\frac{x_1}{z_1}\Big) 
\, f_{b/H_2}\Big(\frac{x_2}{z_2}\Big) \,,
\en
where $f_{a/H}(\xi)$ with $\xi_i = x_i/z_i$
is the PDF describing the collinear $\xi$ 
distribution of partons of type $a$ in a hadron $H$.

As we know the DY angular coefficients $A_0$ and $A_2$ at NLO 
are related by the LT identity $A_0=A_2$. 
For the first time this identity was discovered 
in the naive parton model~\cite{Lam:1978zr} and then confirmed 
in pQCD at order ${\cal O}(\alpha_s)$ based on the collinear factorization 
(see refs.~\cite{Collins:1984kg,Boer:2006eq,Berger:2007jw}). Violation of the LT identity was experimentally observed and numerically simulated using different MC generators, violation starts at computation of NNLO contribution 
to the DY cross-section~\cite{ATLAS:2016rnf,Gauld:2017tww}.

\subsection{Geometrical method}

\begin{figure}[t]
    \centering
    \includegraphics[height=6cm,trim={0cm 1.8cm 0cm 0cm},clip]
    {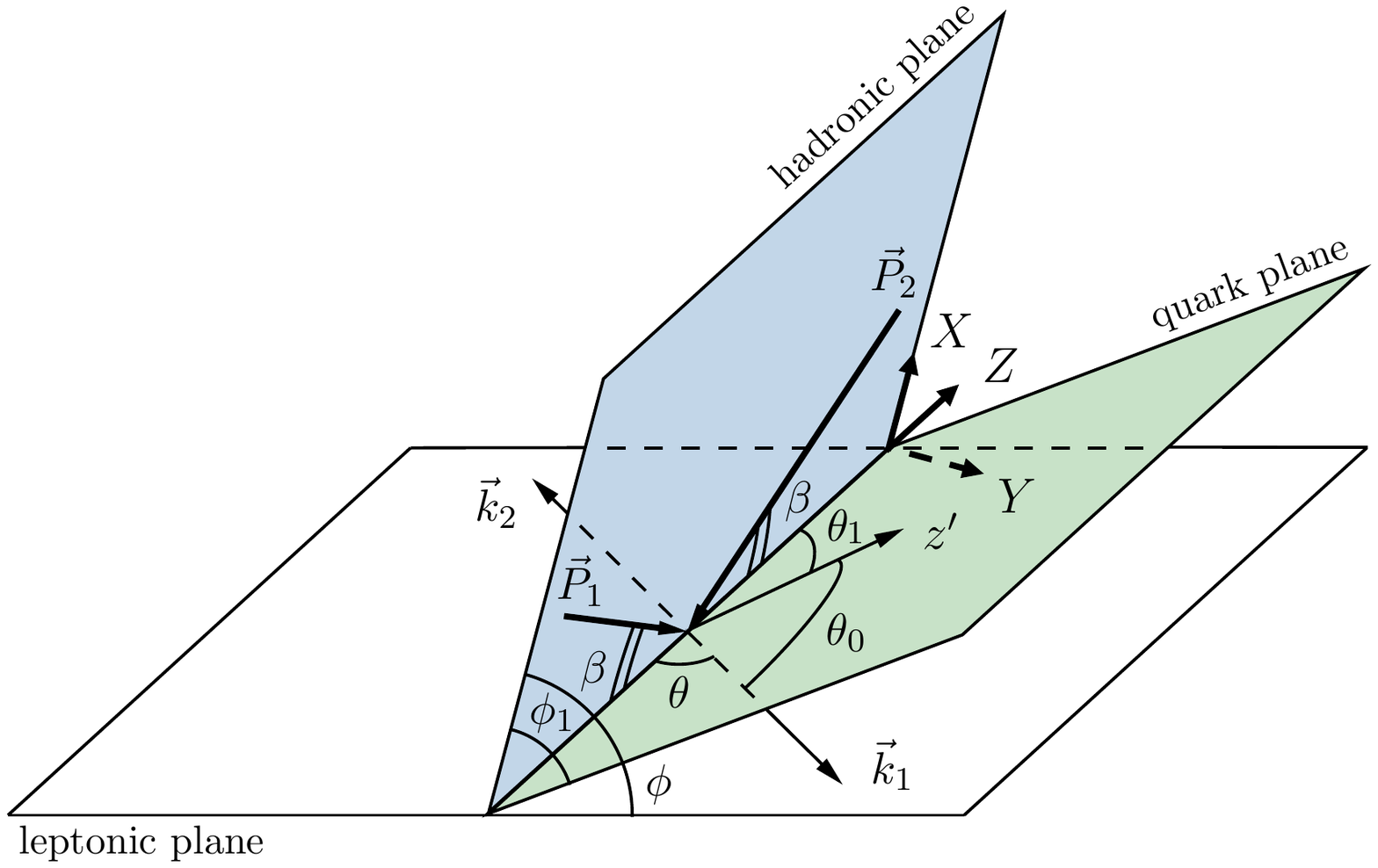}
    \caption{DY process in the CS frame. 
    The hadron, quark, and lepton planes are colored
    by blue, green, and white color, respectively.}
    \label{cs_frame}
\end{figure}

In refs.~\cite{Argyres:1982kg,Peng:2015spa,Chang:2017kuv} it was proposed and developed the geometrical approach for study of the angular coefficients dependence  on small $Q_T$ and rapidity $y$. 
In figure~\ref{cs_frame} we specify the kinematics 
of the DY process in the Collins-Soper (CS) frame. 
CS frame is the rest frame of the virtual ($\gamma^*/Z$) boson or, equivalently, of the lepton pair with special definition of the hadronic plane formed by the spatial momenta of the beam ($\vec{P}_1$) and the target ($\vec{P}_2$) hadrons. 
In particular, in the CS frame the $Z$ axis is fixed as pointing 
in the direction that bisects the angle $\beta$ between the 
three-vectors $\vec{P}_1$ and $-\vec{P}_2$ with $\sin\beta = \rho/\sqrt{1 + \rho^2}$ 
and $\cos\beta = 1/\sqrt{1 + \rho^2}$. In figure~\ref{cs_frame} we define: 
(1) three planes -- hadronic, quark, and leptonic, 
which are specified by pairs of the corresponding three-momenta of 
colliding hadrons, quark and antiquark, lepton and antilepton; 
(2) the angle $\theta_0$, which is the angle between  
lepton $\ell^-$ and quark three-momenta; 
(3) two pairs of polar and azimuthal angles: $(\theta,\phi)$ defining relative orientation of hadronic and leptonic frames and $(\theta_1,\phi_1)$ defining 
relative orientation of hadronic and quark frames. 

The main idea of the geometrical  method~\cite{Argyres:1982kg,Faccioli:2011pn,Peng:2015spa,Chang:2017kuv}  
is to use the "natural" quark-antiquark axis for definition 
of the DY angular lepton distribution. In particular, the latter is 
azimuthally symmetric with respect to this axis and only 
depends on the polar angle $\theta_0$:
\eq\label{qq_sig_ter}
\frac{dN}{d\Omega} &=& \frac{3}{16\pi} \, 
\biggl[1 + \cos^2\theta_0 + a \cos\theta_0\biggr] \,.
\en
Note, the third term in r.h.s. of eq.~(\ref{qq_sig_ter}) 
was added in ref.~\cite{Chang:2017kuv,Lyu:2020nul} for the description 
of the weak gauge boson production. 
Here the coefficient $a$ is the forward-backward (FB) asymmetry 
originating from the parity-violating coupling of quarks and leptons weak gauge boson. 
The angle $\theta_0$ can be expressed through 
the set of angles $\phi$, $\theta_1$, and $\phi_1$ using 
identity~\cite{Argyres:1982kg,Peng:2015spa,Lyu:2020nul}: 
\eq
\label{theta0}
\cos\theta_0 = \cos\theta \cos\theta_1 
+ \sin\theta \sin \theta_1 \cos(\phi -\phi_1) \,.
\en
Using relation~(\ref{theta0}) one can present 
the DY angular distribution in the 
form~\cite{Peng:2015spa,Lyu:2020nul} 
\eq\label{eq5-sf}
\frac{dN}{d\Omega} &=& \frac{3}{16\pi} \, 
\biggl[1+\cos^2\theta +
\frac{\sin^2\theta_1}{2} (1-3\cos^2\theta)\nonumber \\
& +&  \frac{1}{2} \sin 2\theta_1 \cos \phi_1 
\sin 2\theta \cos\phi 
 +  \frac{1}{2} \sin^2\theta_1 \cos 2\phi_1 
\sin^2\theta \cos 2\phi \nonumber \\
& + & a \sin \theta_1 \cos \phi_1 \sin\theta \cos\phi
+ a \cos \theta_1 \cos\theta 
 +  \frac{1}{2} \sin^2\theta_1 \sin 2\phi_1 \sin^2\theta \sin 2\phi
\nonumber \\
& + & \frac{1}{2} \sin 2\theta_1 \sin\phi_1 \sin 2\theta \sin\phi
 +  a \sin\theta_1 \sin\phi_1 \sin\theta \sin\phi\biggr]\,.
\label{eq:eq5Ter}
\en

Matching eqs.~(\ref{eq4-sf}) and~(\ref{eq5-sf}) gives 
for the angular coefficients~\cite{Peng:2015spa,Lyu:2020nul} 
\begin{align}
\label{eq:eq6}
A_0 &= \sin^2\theta_1\,, & 
A_1 &= \frac{1}{2} \sin2\theta_1 \cos \phi_1 \,, & 
A_2 &= \sin^2\theta_1 \cos 2\phi_1 \,, &
A_3 &= a \sin \theta_1 \cos \phi_1 \,, \nonumber\\
A_4 &=  a \cos \theta_1 \,, & 
A_5 &=  \frac{1}{2} \sin^2\theta_1 \sin 2\phi_1 \,, &
A_6 &= \frac{1}{2} \sin 2\theta_1 \sin\phi_1 \,, &
A_7 &=  a \sin\theta_1 \sin\phi_1 \,. 
\end{align}
In case of the quark-antiquark annihilation 
subprocess $q \bar q \to g \gamma(Z,W)$ 
the $\sin\theta_1$ is identified with 
$\sin\beta = \rho/\sqrt{1 + \rho^2}$~\cite{Peng:2015spa,Chang:2017kuv},  
while in case of the quark-gluon Compton scattering 
$q g \to q \gamma(Z,W)$ the $\sin\theta_1$ is approximately equal to 
$\sqrt{5} \rho/\sqrt{1 + 5 \rho^2}$~\cite{Peng:2015spa,Chang:2017kuv}. 

\begin{table}[t!]
    \caption{\label{tab:params}
    Parameters of the geometric method fitted to data from the CMS \cite{CMS:2015cyj}, ATLAS \cite{ATLAS:2016rnf} and LHCb  \cite{LHCb:2022tbc} Collaborations.}
    \label{table_fit_geom}
    \centering
    \vspace{.1cm}
    \begin{tabular}{c|c|c|c|c|c|c}
        \hline
         data & \multicolumn{2}{|c|}{CMS} & \multicolumn{3}{|c|}{ATLAS} & LHCb\\
        \hline
        \diagbox[width=3.8cm]{parameter}{rapidity $|y|$} &  $[0,1]$ & $[1,2.1]$ & $[0,1]$ & $[1,2]$ & $[2,3.5]$ & $[2,4.5]$ \\
        \hline
        $f$ & \multicolumn{2}{|c|}{0.509} & \multicolumn{3}{|c|}{0.446} & 0.619\\
        \hline
        $r_1$ & 0.024 & 0.112 & 0.028 & 0.130 & --- & 0.267\\
        \hline
        $r_3$ & 0.015 & 0.034 & 0.000 & 0.032 & 0.093 & 0.099\\
        \hline
        $r_4$ & 0.018 & 0.073 & 0.021 & 0.063 & 0.136 & 0.143\\
        \hline
    \end{tabular}
\end{table}

In the geometrical approach all angular T-even coefficients 
$A_{i;\text{g}}$ are given by linear combinations of 
$q\bar{q}$ and $qg$ contributions using mixing parameter 
$f$~\cite{Chang:2017kuv}:
\eq
A_{0;\text{g}} &=& f \, A_{0;\text{g}}^{q\bar{q}} + (1-f)\,A_{0;\text{g}}^{qg} 
= f\,\frac{\rho^2}{1 + \rho^2} 
+ (1-f)\,\frac{5 \rho^2}{1 + \rho^2} 
\,, \nonumber\\
A_{1;\text{g}} &=& f \, A_{1;\text{g}}^{q\bar{q}} + (1-f)\,A_{1;\text{g}}^{qg}=r_1\,\bigg[f\,\frac{\rho}{1 + \rho^2} + (1-f)\,\frac{\sqrt{5}\,\rho}
{1 + 5\, \rho^2}\bigg]
\,, \nonumber\\
A_{2;\text{g}} &=& r_2 \, A_{0;\text{g}}
\,, \nonumber\\
A_{3;\text{g}} &=& f \, A_{3;\text{g}}^{q\bar{q}} + (1-f)\,A_{3;\text{g}}^{qg}=r_3\,\bigg[f\,\frac{\rho}{\sqrt{1 + \rho^2}} + (1-f)\,\frac{\sqrt{5}\,\rho}{\sqrt{1 + 5\, \rho^2}}\bigg]
\,, \nonumber\\
A_{4;\text{g}} &=& f \, A_{4;\text{g}}^{q\bar{q}} + (1-f)\,A_{4;\text{g}}^{qg}=r_4\,\bigg[f\,\frac{1}{\sqrt{1 + \rho^2}} 
+ (1-f)\,\frac{1}{\sqrt{1 + 5\,\rho^2}}\bigg]\,,
\en
where $r_i$ with $i=1,2,3,4$ are the reduction factors, which play certain 
role in the angular coefficients phenomenology. In particular, the 
$r_1$ is related to the rotation of the hadronic plane relative to 
the quark plane, $r_2$ parametrizes a violation of the LT identity $A_0=A_2$ 
(we remind that LT relation is violated starting from NNLO ($\alpha_s^2$)~\cite{Mirkes:1992hu}), $r_3$ and $r_4$ are related to the 
forward-backward asymmetry. The parameters  $r_1$, $r_3$, and $r_4$ are fitted accordingly using different experimental datasets at specific values of total energy $\sqrt{s}$ and rapidity $y$. After fitting free parameters the 
geometrical method can predict a behavior of angular coefficients in some 
interval of $Q_T$. We test the geometrical method by comparison with the results for angular coefficients computed by us in the pQCD at NLO  based on the collinear factorization~\cite{Lyubovitskij:2024jlb}. 
We make statistical analysis using currently available data from 
CMS~\cite{CMS:2015cyj}, ATLAS~\cite{ATLAS:2016rnf},  
and LHCb~\cite{LHCb:2022tbc} Collaborations. 
Parameters of the geometrical method fitted to different sets of 
the LHC data are shown in Table~\ref{tab:params}. 
Note, in the case of comparisons with LHCb data~\cite{LHCb:2022tbc}  
the rapidity region is strictly forward.

\subsection{DY partonic structure functions in pQCD}   

\begin{figure}[t]
    \centering
    \includegraphics[height=4.5cm,trim={0cm 2cm 0cm 2cm},clip]{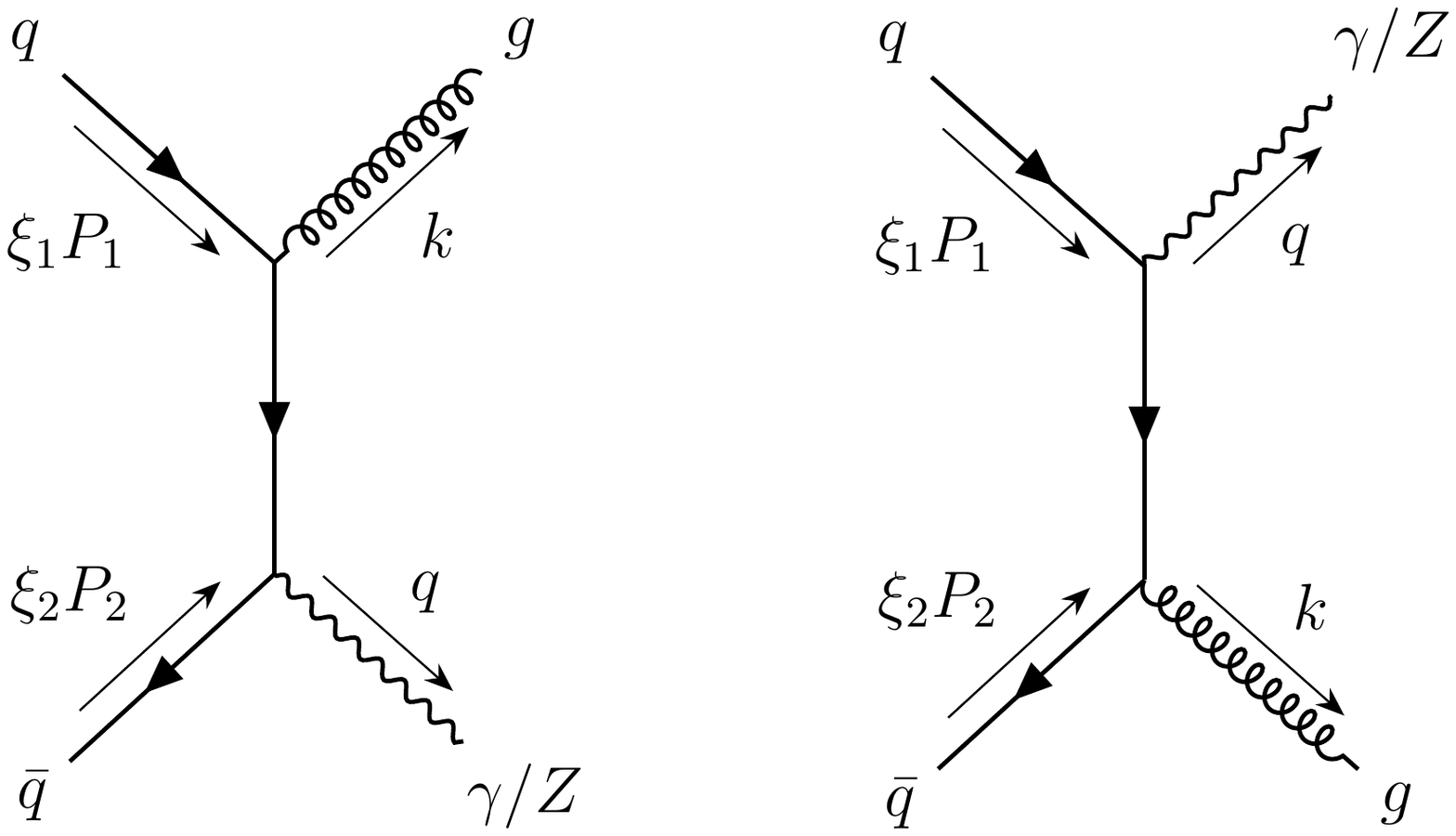}
    \caption{Partonic-level quark-antiquark annihilation diagrams that contribute to the DY cross section at order $\mathcal{O}(\alpha_s)$.}
    \label{qq_diag}

\vspace*{.5cm}

    \centering
    \hspace*{-3.2cm}
    \includegraphics[height=4.5cm,trim={-3cm 2.6cm 8cm 2cm},clip]{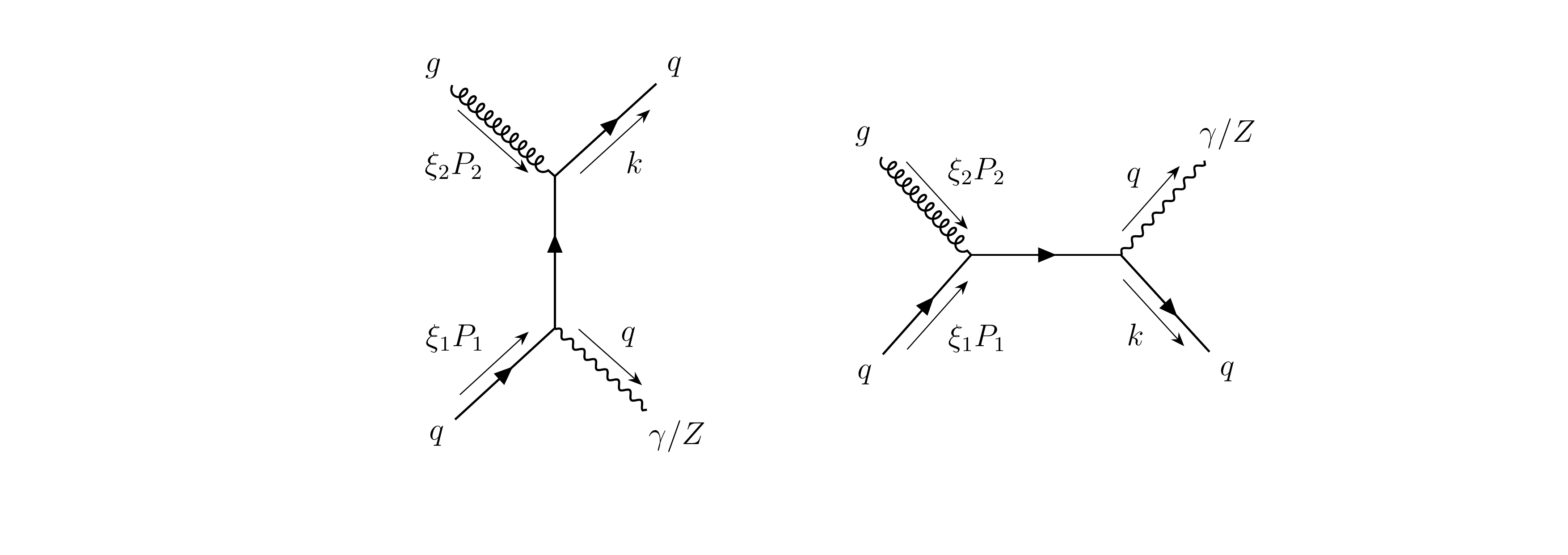}
    \caption{Partonic-level quark-gluon Compton scattering diagrams that contribute to the\\ DY cross section at order $\mathcal{O}(\alpha_s)$.}
    \label{qg_diag}
\end{figure}

Making analysis we use our analytical results for the DY 
angular coefficients computed in ref.~\cite{Lyubovitskij:2024jlb} 
using collinear pQCD in the first order of $\alpha_s$ (NLO), i.e. 
at the same order of accuracy as it was done in the geometrical approach 
and taking into account of partonic subprocesses of quark-antiquark 
annihilation (see figure~\ref{qq_diag}) 
and Compton quark (antiquark)-gluon scattering (see figure~\ref{qg_diag}). 

Angular coefficients are related to partonic-level structure functions 
with eqs.~(\ref{A_Wrelations}) and~(\ref{factorization}). 
Next we specify partonic-level structure functions occurring 
at NLO~\cite{Lyubovitskij:2024jlb,Berger:2007jw,Boer:2006eq} 
\eq
\label{phase_space} 
w_j^{ab}(z_1,z_2,\rho^2)\,&=&\,\tilde{w}_j^{ab}(z_1,z_2,\rho^2)
\,\delta\big((\hat{s}+\hat{t}+\hat{u}-Q^2)/\hat{s}\big) \\
&=&\,\tilde{w}_j^{ab}(z_1,z_2,\rho^2)
\,\delta\left((1-z_1)(1-z_2)
-\frac{\rho^2 \,  z_1 z_2}{1+\rho^2}\right) \nonumber
\en
where we use introduced before the variables $z_1=x_1/\xi_1$ and $z_2=x_2/\xi_2$. 

Here $\hat{s}$, $\hat{t}$, and $\hat{u}$ are parton-level Mandelstam variables are expressed and obey the identities as
\eq 
& &\hat{s} = \frac{Q^2 + Q_T^2}{z_1 z_2} = 
\frac{Q_T^2}{(1-z_1)(1-z_2)} \,,\nonumber\\
& &\hat{t} = Q^2 - \frac{Q^2 + Q_T^2}{z_1} 
= - \frac{Q_T^2}{1-z_2} \,,\nonumber\\ 
& &\hat{u} = Q^2 - \frac{Q^2 + Q_T^2}{z_2} 
= - \frac{Q_T^2}{1-z_1} \,,\\
& &   \frac{\hat{t}}{\hat{s}} = z_1 - 1\,,
\quad \frac{\hat{u}}{\hat{s}} = z_2 - 1\,,
\quad \frac{Q^2}{\hat{s}} = z_1 + z_2 - 1
\,,\nonumber
\en
The parton-level helicity structure functions from $q\bar{q}$ annihilation 
subprocess have a form
\eq
\label{qq_func1z1z2}
\tilde{w}_T^{q\bar q} &=& g_{q\bar q; 1} 
\, \frac{1}{\rho^2} \, 
\biggl(1 + \frac{\rho^2}{2} \biggr) 
\, \frac{z_1^2 + z_2^2}{z_1 z_2}  
\,,  \nn\\[1mm] 
\tilde{w}_L^{q\bar q} &=& 2 \, \omega_{\Delta\Delta}^{q\bar q}  
\ = \ g_{q\bar q; 1} 
\, \frac{z_1^2 + z_2^2}{z_1 z_2}  
\,, \nn\\[1mm] 
\tilde{w}_\Delta^{q\bar q} &=&
g_{q\bar q; 1} 
\,  \frac{1}{\rho} \,
\, \frac{z_1^2 - z_2^2}{z_1 z_2}  
\,,  \nn\\[1mm] 
\tilde{w}_{T_P}^{q\bar q} &=&
g_{q\bar q; 2} 
\,  \frac{\sqrt{1+\rho^2}}{\rho^2}  
\, \frac{z_1^2 + z_2^2}{z_1 z_2}
\,,  \nn\\[1mm] 
\tilde{w}_{\nabla_P}^{q\bar q}                                                 
&=& g_{q\bar q; 2}
\,  \frac{\sqrt{1+\rho^2}}{\rho}  
\, \frac{z_1^2 - z_2^2}{z_1 z_2}  \,. 
\en

For the $qg$ subprocess we have
\eq
\label{qg_func1z1z2}
\tilde{w}_T^{q g} &=& g_{q g; 1} \,  \frac{1}{\rho^2} \,
\frac{1-z_2}{z_1 z_2} \,
\biggl(z_2^2 + (1 - z_1 z_2)^2  + \rho^2
\Big(1 - \frac{z_1^2}{2} - z_1 z_2 (z_1 + z_2)\Big)\biggr)  
\,,  \nn\\[1mm] 
\tilde{w}_L^{q g} &=& 2 \, \omega_{\Delta\Delta}^{q g}
\ = \ g_{q g; 1} \, 
\frac{1-z_2}{z_1 z_2} \ \Big(z_2^2 + (z_1 + z_2)^2\Big) 
\,, \nn\\
\tilde{w}_\Delta^{q g} &=&
g_{q g; 1} \, \frac{1}{\rho} \,
\frac{1-z_2}{z_1 z_2} \ \Big(z_1^2 - 2 z_2^2\Big) 
\,,  \nn\\[1mm] 
\tilde{w}_{T_P}^{q g} &=&
g_{q g; 2} \,  \frac{\sqrt{1+\rho^2}}{\rho^2}  
\, \frac{1-z_2}{z_1 z_2}
\ \Big(z_2^2 + (1 - z_2)^2 - (1 - z_1)^2\Big)
\,,  \nn\\[1mm] 
\tilde{w}_{\nabla_P}^{q g}                                                 
&=& g_{q g; 2} \,  \frac{\sqrt{1+\rho^2}}{\rho}  
\, \frac{1-z_2}{z_1 z_2} \ 
\Big(1- 2 z_2^2 - (1-z_1)^2 + 2 z_2 (1 - z_1) \Big) \,.
\en
Details on how these distributions are changing at changing types of partons are discussed in ref.~\cite{Lyubovitskij:2024jlb}. 
In equations~(\ref{qq_func1z1z2}) and~(\ref{qg_func1z1z2})
we introduce couplings $g_{q\bar q; i}$ and $g_{qg; i}$ 
defined as 
\eq
g_{q\bar q; i} &=& (8 \, \pi^2 \e_q^2 \, \alpha_s) \, 
C_{q\bar q} \, g_{{\rm EW}; i}^{Z\gamma/W} \,,\\
g_{qg; i} &= & (8 \, \pi^2 \e_q^2 \, \alpha_s) \,
C_{qg} \,  g_{{\rm EW}; i}^{Z\gamma/W}
\,, 
\en
where $C_{q\bar q} = \frac{4}{9}$, 
$C_{qg}  = \frac{1}{6}$,  $C_F =\frac{4}{3}$, 
$C_A = 3 $ and $T_F = \frac{1}{2}$, $e_q$ is charge of quarks.
For calculation of the $T$-even structure functions 
we need the following combinations of weak couplings 
and Breit-Wigner propagator of weak gauge boson: 
\eq
g_{\rm EW; 1}^{Z\gamma} &=& 1 + 2 \, g^V_{Zq} \, g^V_{Zl} \, 
\mathrm{Re}[D_Z(Q^2)] + \Big((g_{Zq}^{V})^2 + (g_{Zq}^{A})^2\Big)  \Big((g_{Z\ell}^{V})^2 + (g_{Z\ell}^{A})^2\Big) \,|D_Z(Q^2)|^2\,,
\nonumber\\[2mm] 
g^{Z\gamma}_{\rm EW; 2}&=&4 \, g_{Zq}^{A} \, \Big[
2 \, g_{Zq}^{V}  \, \Big(g_{Z\ell}^{A} \, g_{Z\ell}^{V}\Big) \, 
|D_Z(Q^2)|^2 + g_{Z\ell}^{A} \, \mathrm{Re}[D_Z(Q^2)]\Big] \,,
\en
where 
\eq
g_{Z\ell}^{V} &=& - \frac{1 - 4 \sin^2\theta_W}{2 \sin 2\theta_W}
\,,
\quad \hspace*{.1cm} 
g_{Z\ell}^{A} \, = \, - \frac{1}{2 \sin 2\theta_W}
\,,
\nonumber\\[2mm] 
g_{Zu}^{V} &=& \frac{1 - 8/3 \sin^2\theta_W}{2 e_q \sin 2\theta_W}
\,,
\quad 
g_{Zd}^{V} \, = \, - \frac{1 - 4/3 \sin^2\theta_W}{2 e_q \sin 2\theta_W}
\,,
\nonumber\\[2mm] 
g_{Zu}^{A} &=&   \frac{1}{2 e_q \sin 2\theta_W}
\,,
\quad \hspace*{.6cm} 
g_{Zd}^{A} \, = \, - \frac{1}{2 e_q \sin 2\theta_W}    
\en
and 
\eq
\mathrm{Re}[D_G(Q^2)] &=& 
\frac{(M_G^2-Q^2) Q^2}{(M_G^2-Q^2)^2 + M_G^2 \Gamma_G^2} \,, 
\nonumber\\[2mm] 
\mathrm{Im}[D_G(Q^2)] &=& 
\frac{M_G \Gamma_G Q^2}{(M_G^2-Q^2)^2 + M_G^2 \Gamma_G^2} \,. 
\en
Here Weinberg angle $\theta_W$, masses $M_{G}$, 
and total widths $\Gamma_{G}$ of the weak gauge bosons are 
taken from the Particle Data Group~\cite{ParticleDataGroup:2024cfk} 
$\sin^2\theta_W = 0.23129 \pm 0.00004$, 
$M_{W^\pm} = 80.3692 \pm 0.0133$~GeV,
$M_{Z} = 91.1880 \pm 0.0020$~GeV,
$\Gamma_{W^\pm} = 2.085 \pm 0.042$~GeV,
$\Gamma_{Z} = 2.4955 \pm 0.0023$~GeV.
To perform numerical calculations, 
we utilize the LHAPDF library~\cite{Buckley:2014ana}, 
specifically employing the CT18LO parametrization for PDFs from the CTEQ-TEA Collaboration~\cite{Yan:2022pzl}. Which order of PDF need to use for computation of different order of pertubative part calculation is discussed 
in~\cite{Sherstnev:2007nd}. 

\section{Comparison of the pQCD and geometric approach 
predictions with data}
\label{Sec3_comparison}

In this section, we compare the predictions for the DY angular coefficients obtained in two theoretical approaches (pQCD approach 
at NLO~\cite{Lyubovitskij:2024jlb}) and the geometric approach~\cite{Peng:2015spa,Chang:2017kuv}) with available data 
from the LHC experiments (CMS~\cite{CMS:2015cyj}, ATLAS~\cite{ATLAS:2016rnf}, and LHCb~\cite{LHCb:2022tbc}).

Let us specify how we proceed: \\
(1) In our analysis 
we use unregularized LHC data for different ranges of rapidity 
and do not use regularized data from ATALS Collaboration;\\
(2) For every range of rapidity we take into account 
results of the geometrical method specified by 
an appropriate fit of free parameters (see the list of fitted parameters 
in table~\ref{table_fit_geom}); \\
(3) The geometrical method is based on fitting of 
the set parameters $f, r_i$, which occur in 
the expressions for angular coefficients contributed by 
the sum $q\bar{q}$ and $qg$ subprocesses; \\
(4) In our analysis we also consider total sum of  subprocesses 
$q\bar{q}$, $\bar{q}q$ and $qg$, $gq$, $\bar{q}g$, $g\bar{q}$
as separate contributions of the $q\bar{q}$ and $qg$ subprocesses; \\
(5) Since experimental measurements were not performed at 
fixed value of $Q$ but rather within a $Z$-boson invariant mass window 
($Q \in [81, 101]$ GeV for CMS, $Q \in [80, 100]$ GeV for ATLAS, and 
$Q \in [75, 105]$ GeV for the LHCb Collaboration), we simulate this  
kinematic region in both approaches to ensure a correct comparison with data.  For the geometric method the central curve was fitted at $Q = 90$ GeV; \\ 
(6) In our plots we present the bands bounded by the two curves due to 
a variation of $Q$ from its minimal to maximal experimental values; \\
(7) In contrast, for our pQCD calculations we perform a numerical simulation by randomly sampling from a normal distribution within the same $Q$ range as in the corresponding experiment. For each $Q_T$ value, this process generates a band that accounts for both the simulation procedure and numerical integration uncertainties. For clarity, in all subsequent figures, we omit the central curve and retained only the band bounded by the aforementioned features.

\subsection{Angular coefficient \texorpdfstring{$A_0$}{A0}} 
\label{sec:A0} 
\begin{figure}[t!!]
\centering
\includegraphics[width=0.45\textwidth]{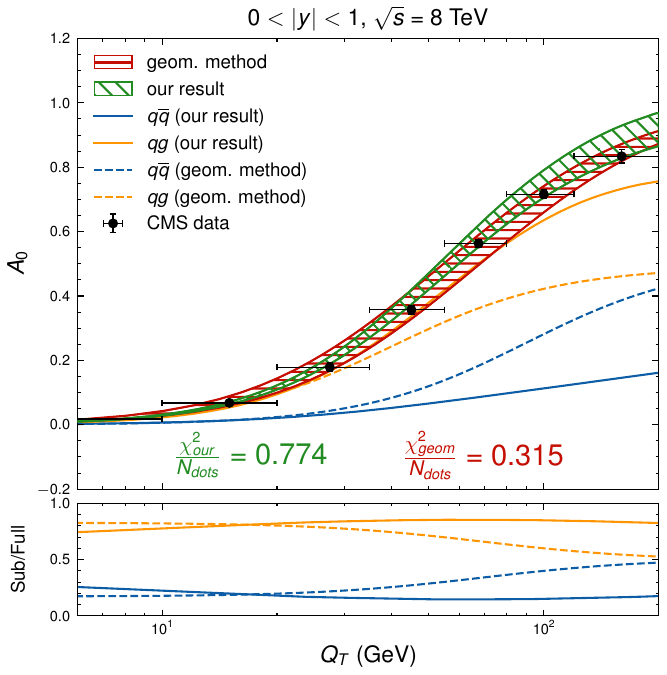}
\includegraphics[width=0.45\textwidth]{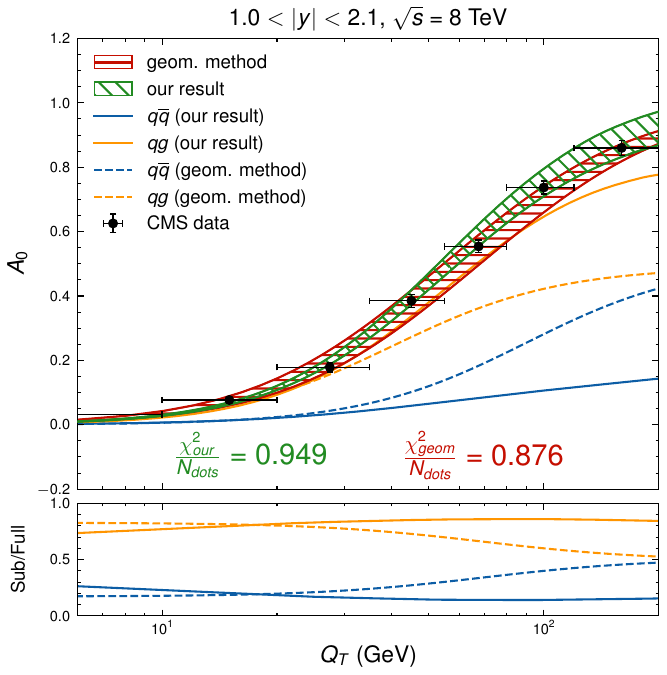}
\includegraphics[width=0.45\textwidth]{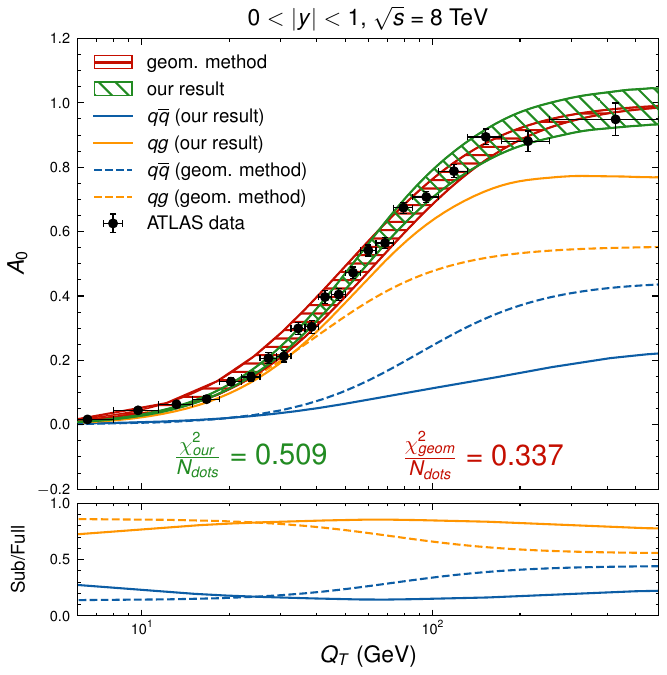}
\includegraphics[width=0.45\textwidth]{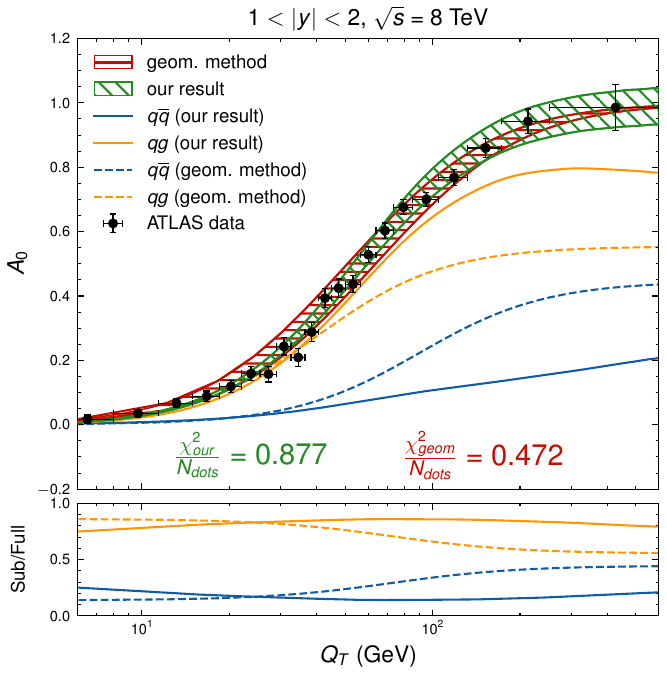}
\includegraphics[width=0.45\textwidth]{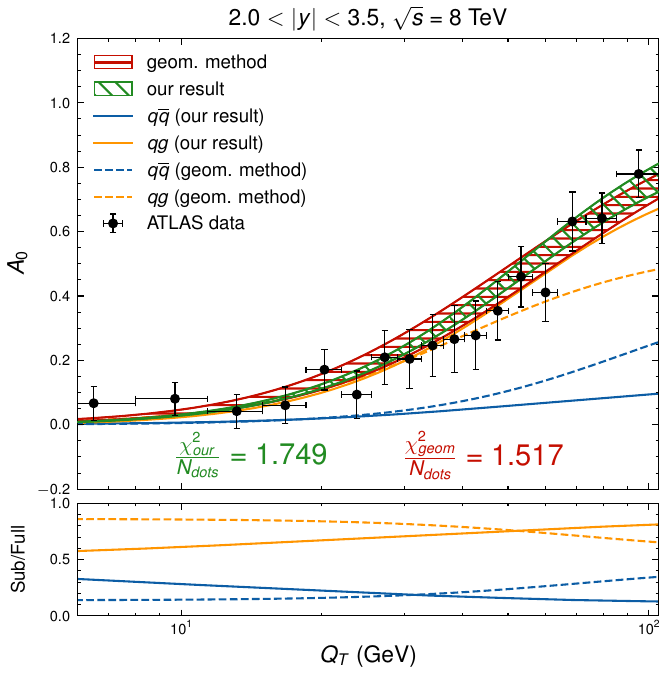}
\includegraphics[width=0.45\textwidth]{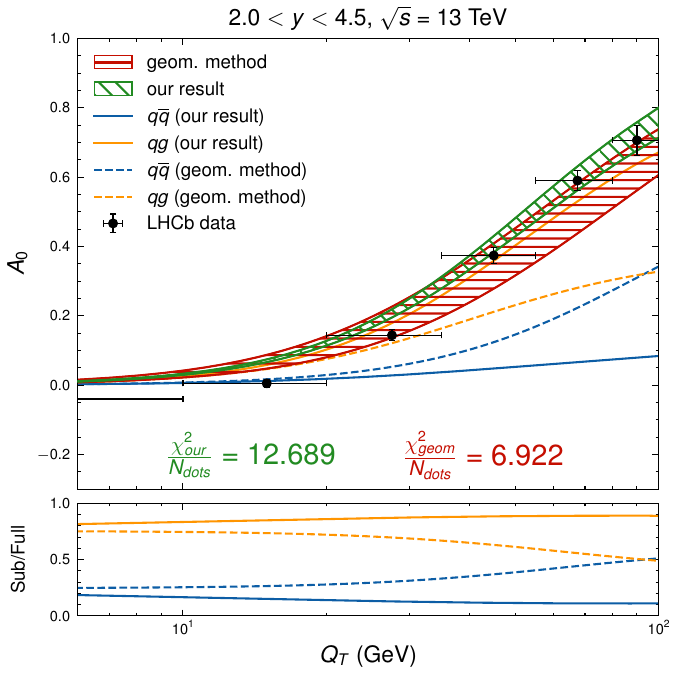}
\caption{\label{fig:atlaslhcba0}
Angular coefficient $A_0$: 
comparison of theoretical results (pQCD and geometric method)  
with CMS~\cite{CMS:2015cyj}, ATLAS~\cite{ATLAS:2016rnf}, 
and LHCb~\cite{LHCb:2022tbc} data 
at $\sqrt{s}=8$~TeV in five rapidity regions 
$0<|y|<1$ and $1<|y|<2.1$, $0<|y|<1$, $1<|y|<2$, and $2<|y|<3.5$ and 
at $\sqrt{s}=13$~TeV in rapidity region 
$2<y<4.5$.}
\label{CMS_Atlas_LHCb_A0_fig}
\end{figure}

When we include the contribution of all perturbative 
subprocesses $q\bar{q}$, $\bar{q}q$, $qg$, $gq$, 
$\bar{q}g$, and $g\bar{q}$ to the angular coefficient $A_0$, 
the latter is symmetric under interchange 
$z_1 \leftrightarrow z_2$ or $y \leftrightarrow - y$. 
In addition, $A_0$ has stable behavior at change of the rapidity 
range. In figure~\ref{CMS_Atlas_LHCb_A0_fig}
we perform a comparison of theoretical results 
(pQCD at NLO and geometrical method) with available 
data from the LHC experiments (CMS data~\cite{CMS:2015cyj}, ATLAS~\cite{ATLAS:2016rnf}, and LHCb~\cite{LHCb:2022tbc}). 
Moreover, figure~\ref{CMS_Atlas_LHCb_A0_fig}
show the results of the reduced $\chi^2$ test for statistical consistency between theoretical predictions and experimental measurements applied to both theoretical approaches 
(pQCD at NLO and geometric method). One can see that
the predictions of both theoretical frameworks demonstrate
reasonable agreement with ATLAS and CMS data in the rapidity 
range $|y|<2$, although the geometric method curve displays 
overfitting behavior in most scenarios. 
Note, for the ATLAS data in the rapidity range $2<|y|<3.5$  agreement is slightly worse for both theoretical frameworks. Furthermore, both pQCD and geometric curves exhibit poor agreement with LHCb data points in the rapidity range $2 < y < 4.5$. It is worth noting that in the LHCb data case apparent agreement is observed, but the statistical results remain poor. This discrepancy arises due to the limited number of data points and the small experimental uncertainties associated 
with measurements in the low-$Q_T$ region. 

We also show the partial contributions of different subprocesses to 
the $Q_T$ dependence of the angular coefficient $A_0$. 
In pQCD the $qg$ subprocess gives the main contribution for the 
LHC energies due to large-scale dominance of the gluon PDF. 
We should note that contributions of the partonic subprocesses 
in the geometrical method are in some inconsistency with the ones 
produced by the pQCD at NLO. Such a disagreement can be explained 
by specifics of the fitting procedure in the geometrical method. Besides, geometrical method do not take into account the behavior of PDF. It is important for LHC data where dominant small $x$ ranges.

\clearpage 

\begin{figure}[h!]
\centering
\includegraphics[width=0.45\textwidth]{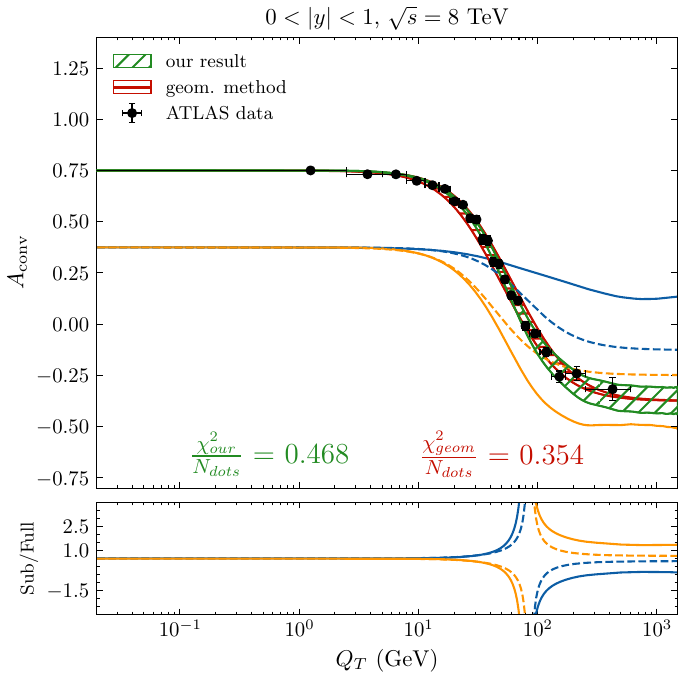}
\includegraphics[width=0.45\textwidth]{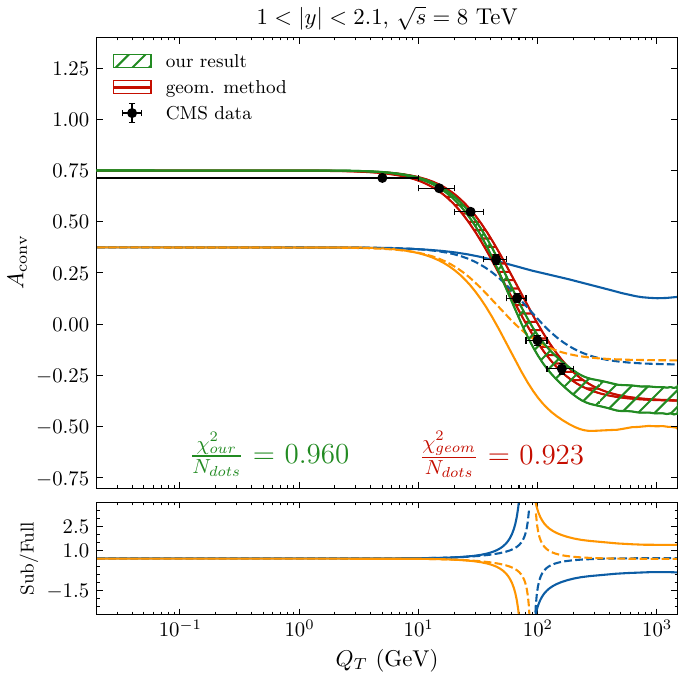}
\caption{\label{fig:Conv}
Convexity asymmetry $A_{conv}$:
comparison of theoretical results (pQCD and geometric method)  
with ATLAS~\cite{ATLAS:2016rnf} and CMS~\cite{CMS:2015cyj} data 
at $\sqrt{s}=8$~TeV in  two rapidity regions $0<|y|<1$ and $1<|y|<2.1$.} 
\end{figure}

In figure~\ref{fig:Conv} for completeness we also show the $Q_T$ behavior of the convexity asymmetry $A_{\rm conv}$~(\ref{Afb_Acon}) proposed in ref.~\cite{Lyubovitskij:2024jlb}. We compare our predictions with ATLAS~\cite{ATLAS:2016rnf} and CMS~\cite{CMS:2015cyj} data 
at $\sqrt{s}=8$ TeV and in rapidity ranges $0<|y|<1$ and $1<|y|<2.1$. 
Here, one needs to note that $A_0$ and $A_{conv}$ at NLO are stable 
to a change of rapidity range.

\subsection{Angular coefficient \texorpdfstring{$A_1$}{A1}}
\label{sec:A1} 

\begin{figure}
\centering
\includegraphics[width=0.45\textwidth]{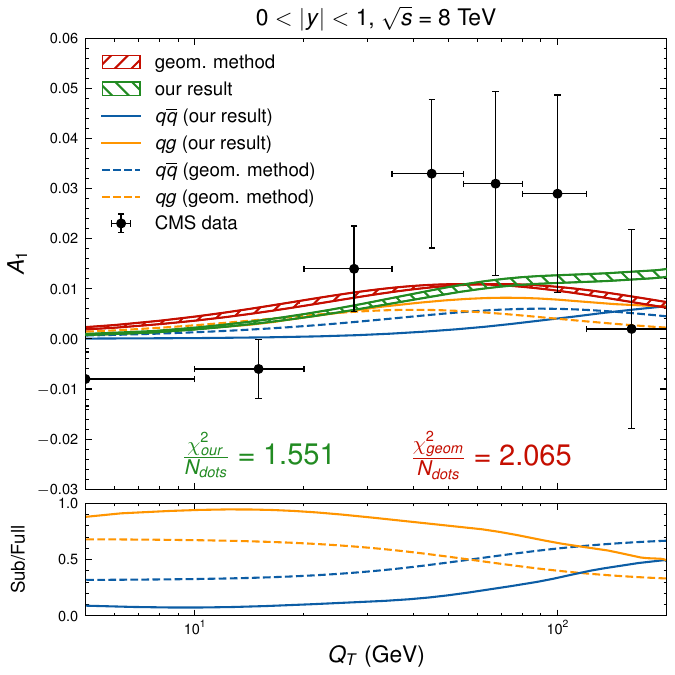}
\includegraphics[width=0.45\textwidth]{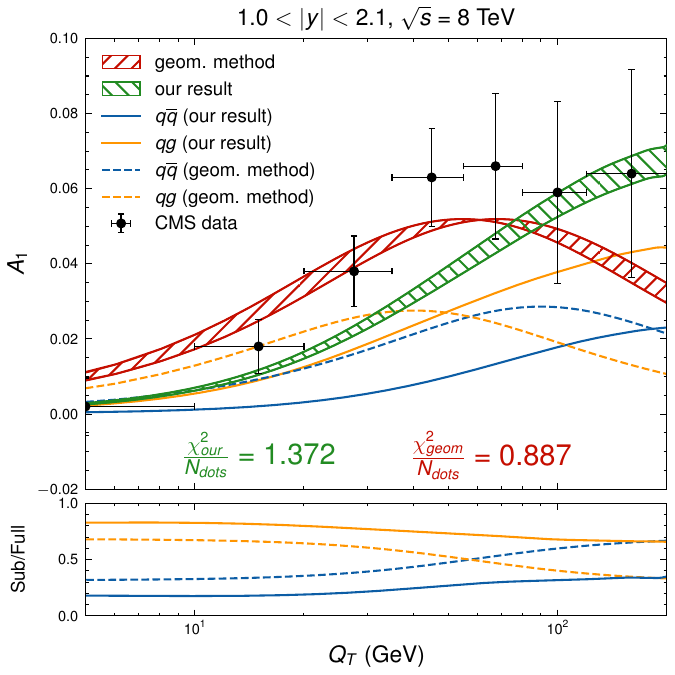}
\includegraphics[width=0.45\textwidth]{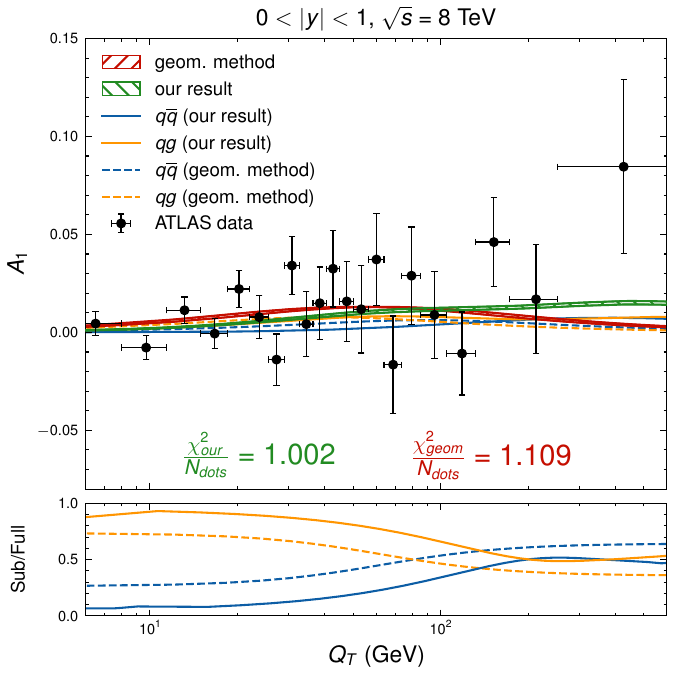}
\includegraphics[width=0.45\textwidth]{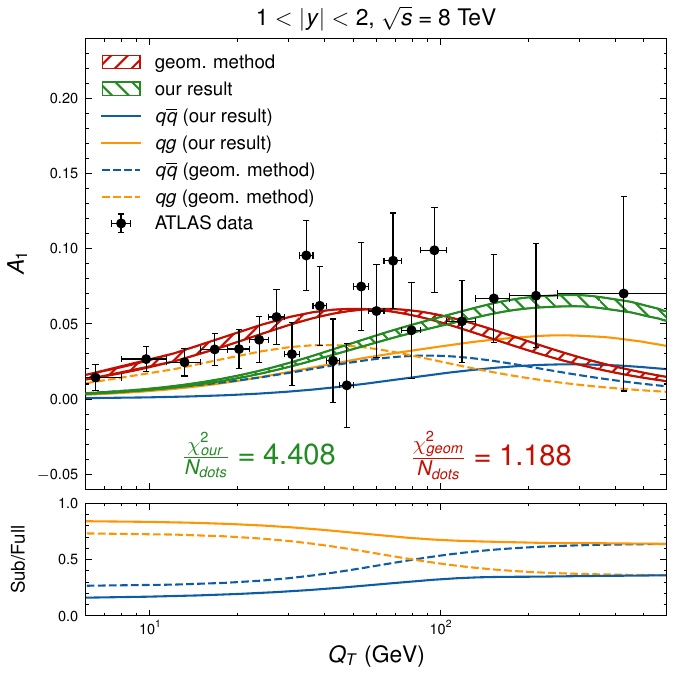}
\includegraphics[width=0.45\textwidth]{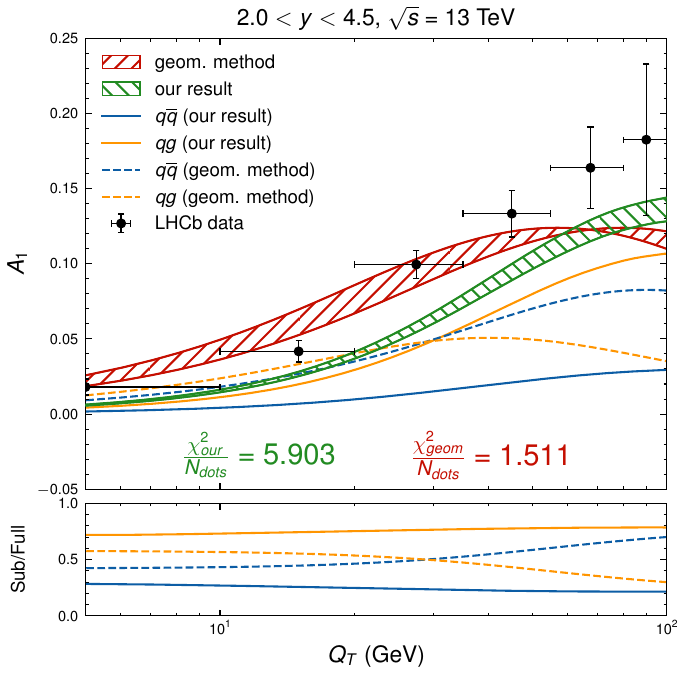}
\caption{\label{fig:cmsatlaslhcb_a1}
Angular coefficient $A_1$:  
comparison of theoretical results (pQCD and geometric method) 
with CMS~\cite{CMS:2015cyj}, ATLAS~\cite{ATLAS:2016rnf}, 
and LHCb~\cite{LHCb:2022tbc} data 
at $\sqrt{s}=8$~TeV in three rapidity regions 
$0<|y|<1$, $1<|y|<2$, and $1<|y|<2.1$ and 
at $\sqrt{s}=13$~TeV in rapidity region 
$2<y<4.5$.}
\end{figure}

Next, in figure~\ref{fig:cmsatlaslhcb_a1} 
we analyze the $Q_T$ dependence of the angular coefficient 
$A_1$ following our strategy in case of the $A_0$. 
This coefficient is connected with single spin-flip hadronic structure function $W_{\Delta}$. 
In the rapidity range $0<|y|<1$ both theoretical approaches have 
pure agreement with CMS data~\cite{CMS:2015cyj}, while they compares much better 
with ATLAS data~\cite{ATLAS:2016rnf}.  
Such a discrepancy with CMS experiment is due to the broader $Q_T$ range, 
which results in a larger number of experimental points for comparison.  
The statistical test against CMS data in the rapidity region $1<|y|< 2.1$ 
demonstrates good agreement of the CMS 
measurements with predictions of both theoretical approaches. 
In contrast, only the geometric method's curve yields good agreement with ATLAS data in the rapidity region $1<|y|<2$.
Finally, there is no agreement of both theoretical approaches 
with LHCb data for rapidity region $2.0<y<4.5$. For pQCD, is shown that behavior at small $Q_T$ is corrected by inclusion of the NNLO correction (see ref.~\cite{ATLAS:2016rnf}).

\begin{figure}
\centering
\includegraphics[width=0.45\textwidth]{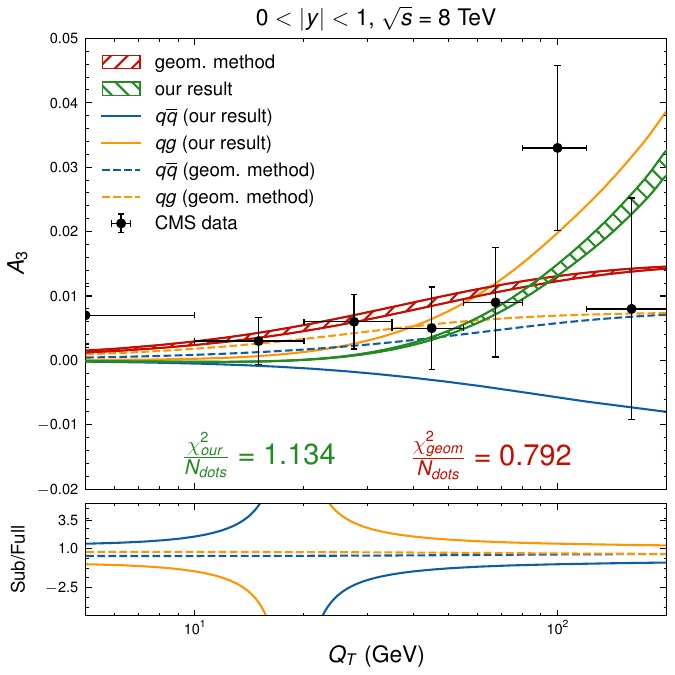}
\includegraphics[width=0.45\textwidth]{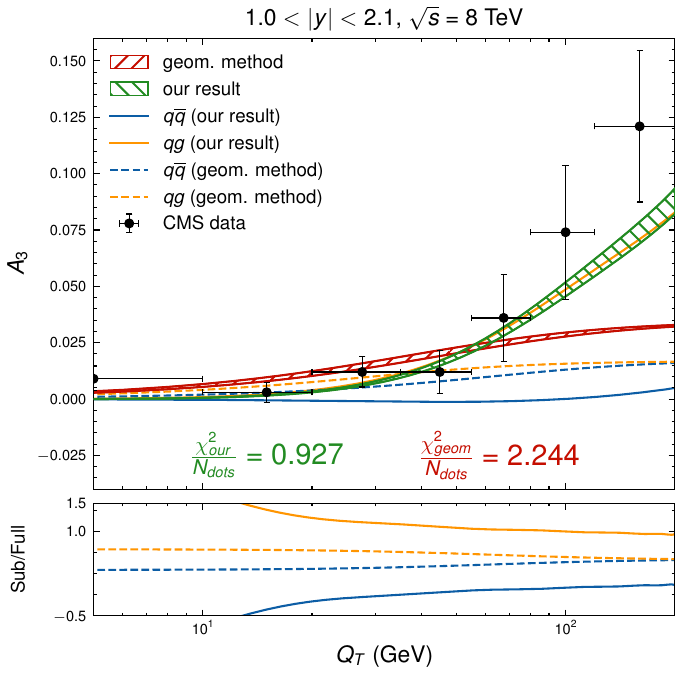}
\includegraphics[width=0.45\textwidth]{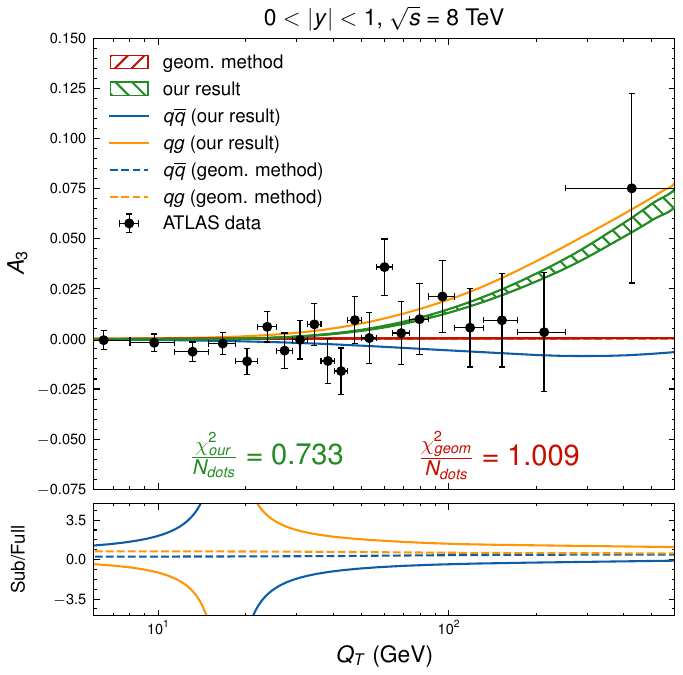}
\includegraphics[width=0.45\textwidth]{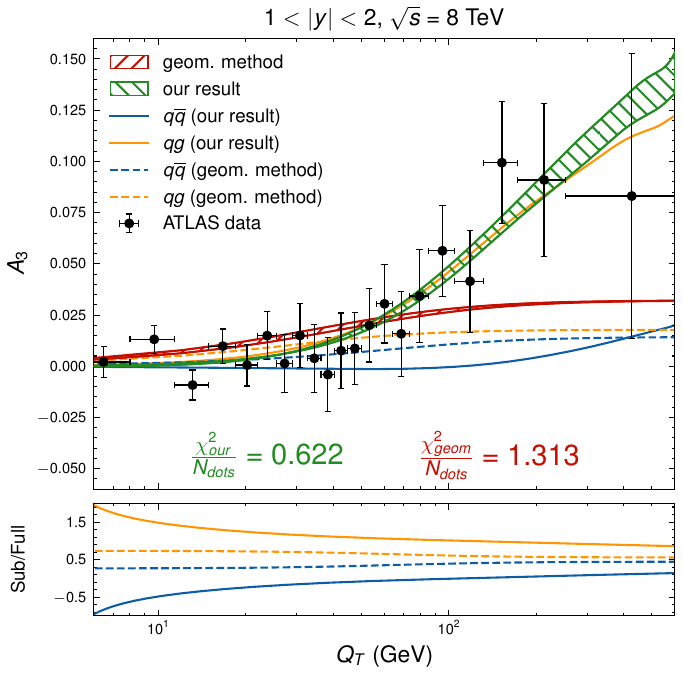}
\includegraphics[width=0.45\textwidth]{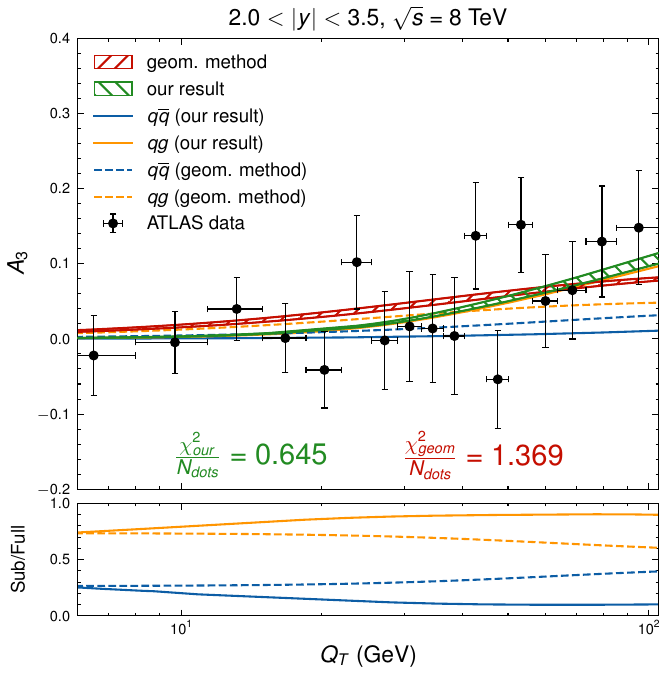}
\includegraphics[width=0.45\textwidth]{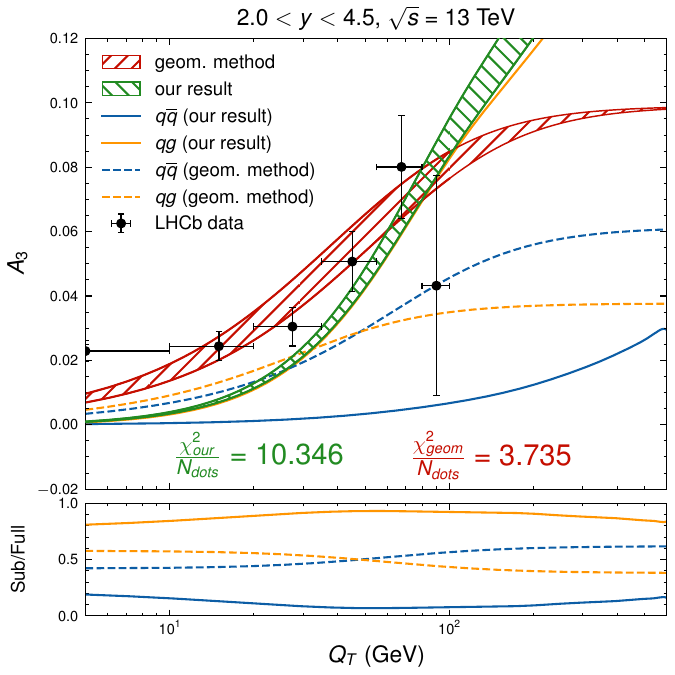}
\caption{\label{fig:cmsatlaslhcb_a3}
Angular coefficient $A_3$:  
comparison of theoretical results (pQCD and geometric method) 
with CMS\cite{CMS:2015cyj}, ATLAS~\cite{ATLAS:2016rnf}, 
and LHCb~\cite{LHCb:2022tbc} data 
at $\sqrt{s}=8$~TeV in four rapidity regions 
$0<|y|<1$, $1<|y|<2$, $1<|y|<2.1$, and $2<|y|<3.5$ 
and at $\sqrt{s}=13$~TeV in rapidity region 
$2<y<4.5$.}
\end{figure}

Regarding the partial contribution of the partonic subprocesses to the total 
$Q_T$ dependence of the coefficient $A_1$ within the pQCD framework, the $qg$ contribution dominates in the whole $Q_T$ range for the most of the rapidity regions. 
However, in the rapidity range $0 < |y| < 1$ the contributions of the $q\bar{q}$ 
and $qg$ subprocesses become comparable for $Q_T>200$~GeV. In the case of the geometric method the picture changes. As $Q_T$ increases, the ratio between the contributions changes and the contribution of the $q\bar{q}$ subprocess becomes dominant, 
rather than the $qg$ one.

\FloatBarrier 

\subsection{Angular coefficient \texorpdfstring{$A_3$}{A3}}
\label{sec:A3}

Analysis of the angular coefficient $A_3$ is shown 
in figure~\ref{fig:cmsatlaslhcb_a3}. 
Except LHCb, pQCD NLO calculations demonstrate excellent 
agreement with ATLAS and CMS measurements for all available 
rapidity ranges. Agreement of the geometric approach 
with ATLAS and CMS data is slightly worse.  
In case of the $A_3$, the $qg$ contribution dominates in both the geometric method and most of the $Q_T$ range within the pQCD framework. The only regions within the pQCD framework,  where the $q\bar{q}$ subprocess exceeds the $qg$ contribution are those where the value of $A_3$ approaches zero. We also note that the geometric method cannot yield negative contributions from individual subprocesses. In contrast, in the rapidity range $|y| < 2.1$, the $q\bar{q}$ subprocess calculated within the pQCD framework contributes negatively to the full $A_3$ coefficient. 

\begin{figure}
\centering
\includegraphics[width=0.45\textwidth]{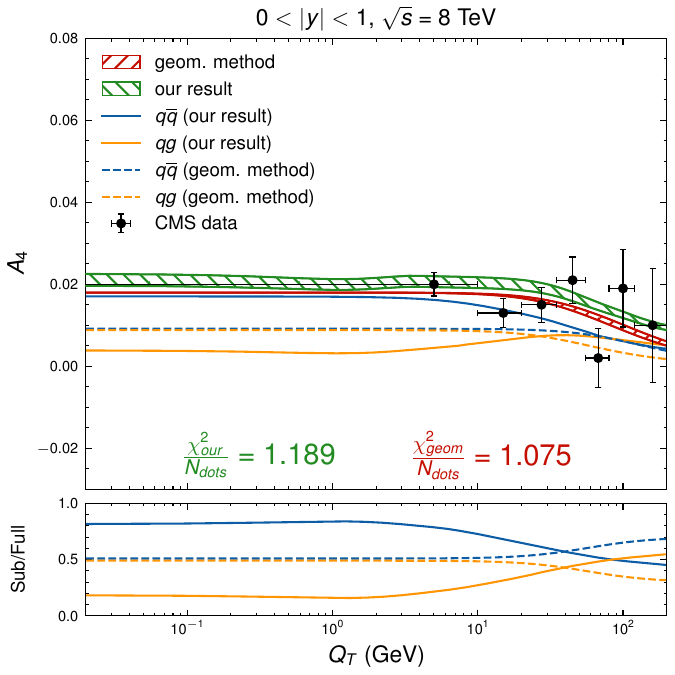}
\includegraphics[width=0.45\textwidth]{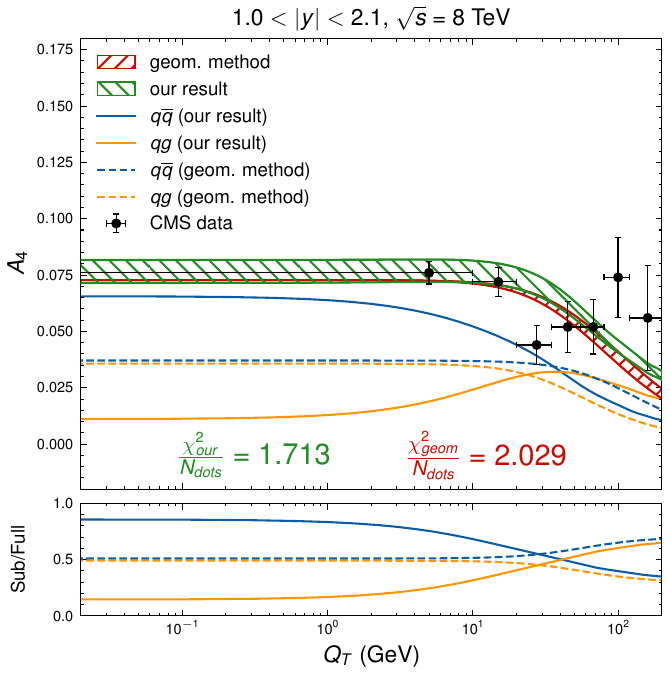}
\includegraphics[width=0.45\textwidth]{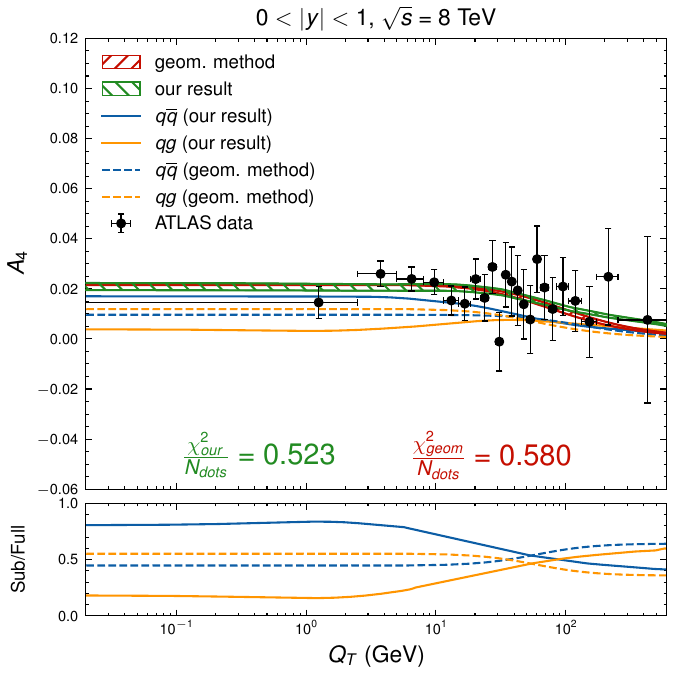}
\includegraphics[width=0.45\textwidth]{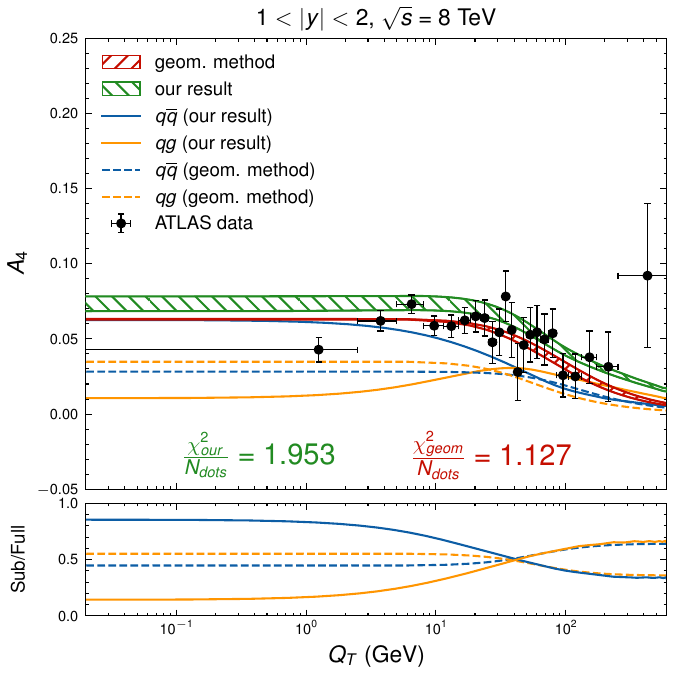}
\includegraphics[width=0.45\textwidth]{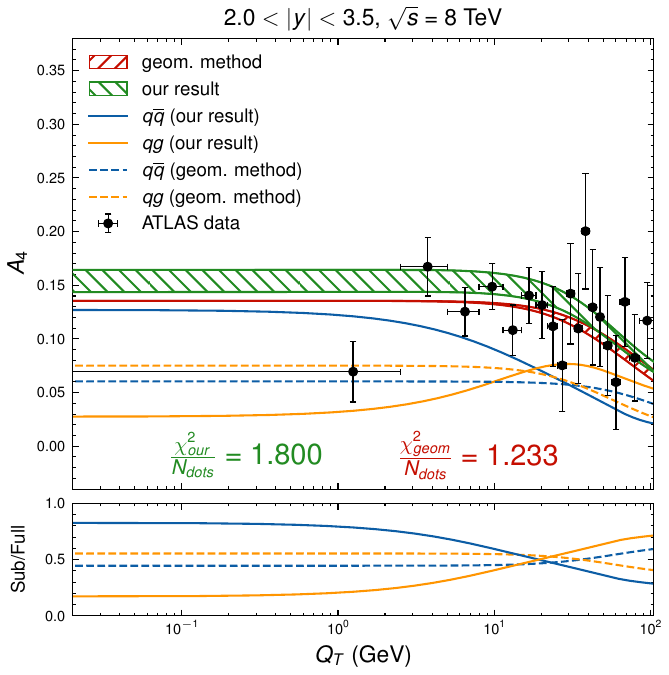}
\includegraphics[width=0.45\textwidth]{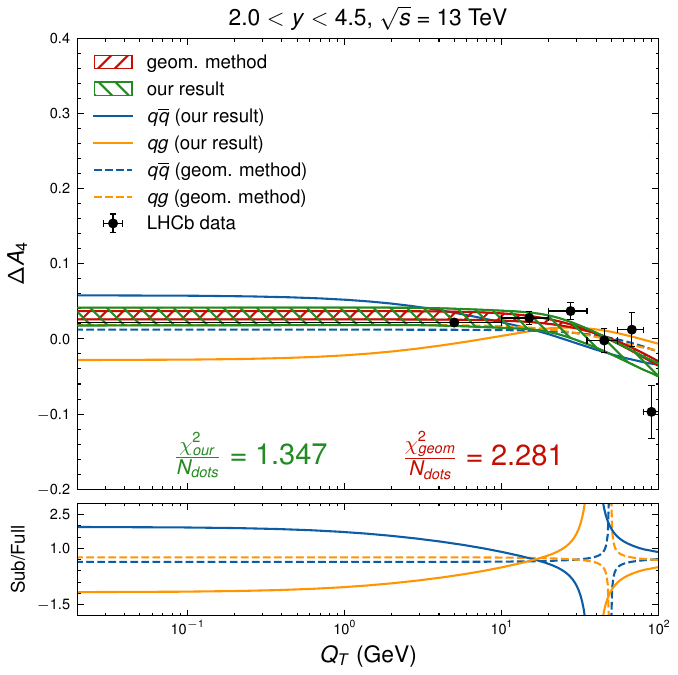}
\caption{\label{fig:cmsatlaslhcb_a4}
Angular coefficient $A_4$: 
comparison of theoretical results (pQCD and geometric method)  
with CMS~\cite{CMS:2015cyj}, ATLAS~\cite{ATLAS:2016rnf}, 
and LHCb~\cite{LHCb:2022tbc} data 
at $\sqrt{s}=8$~TeV in four rapidity regions 
$0<|y|<1$, $1<|y|<2$, $1<|y|<2.1$, and $2<|y|<3.5$ and 
at $\sqrt{s}=13$~TeV in rapidity region 
$2<y<4.5$.}
\end{figure}

\begin{figure}
\centering
\includegraphics[width=0.45\textwidth]{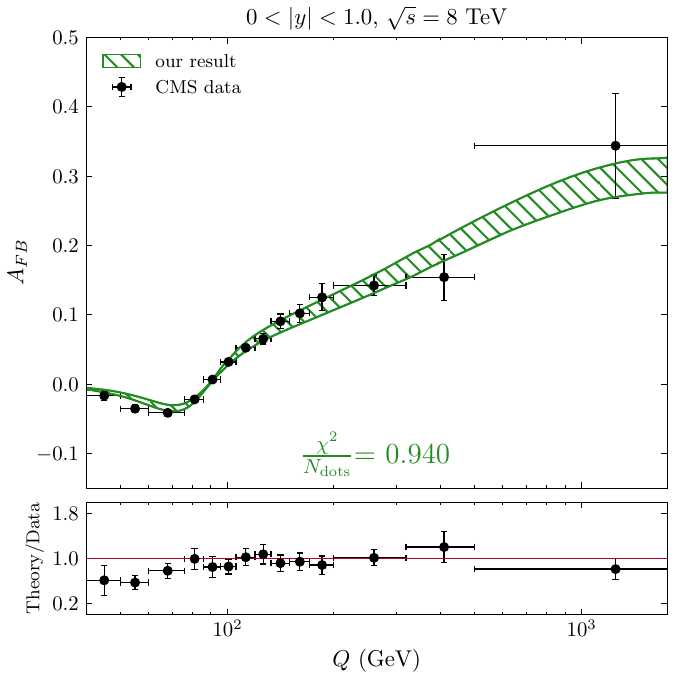}
\includegraphics[width=0.45\textwidth]{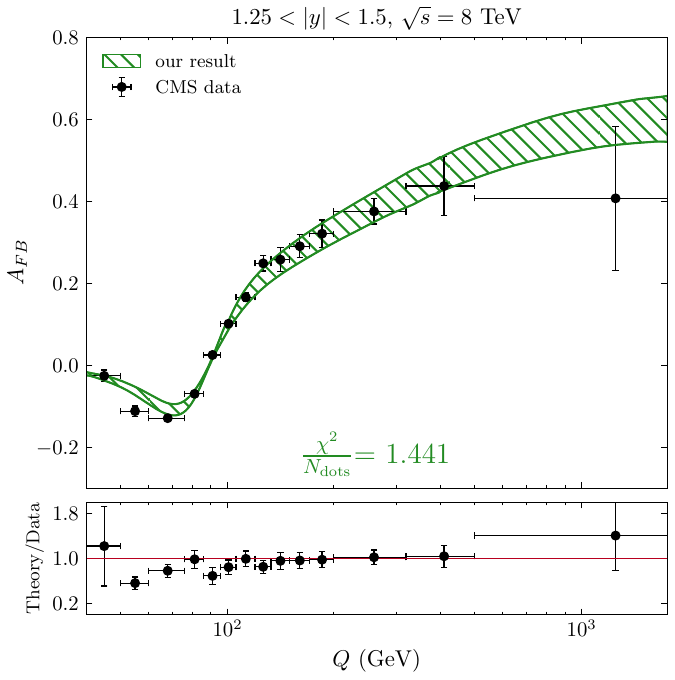}
\caption{\label{fig:FB}
Angular coefficient $A_{FB}$: 
comparison of pQCD at NLO 
with CMS data~\cite{CMS:2016bil} 
at $\sqrt{s}=8$ in two rapidity regions 
$0<|y|<1$ and $1.25<|y|<1.5$.}
\end{figure}
\subsection{Angular coefficient \texorpdfstring{$A_4$}{A4}}
\label{sec:A4}

Finally, we show results for the $A_4$ coefficient 
in figure~\ref{fig:cmsatlaslhcb_a4}. 
One can see that agreement between both theoretical 
frameworks and LHC data is excellent except some 
experimental points with big statistical errors. 
Comparing the individual contributions of the partonic 
subprocesses we see that in the low-$Q_T$ region 
pQCD predicts a dominance of the $q\bar{q}$ subprocess, 
while the $qg$ contribution approaches zero. 
As $Q_T$ increases the contribution of the $qg$ subprocess 
dominates over $q\bar{q}$ one. A similar behavior 
of the individual contributions of the partonic subprocesses 
occurs in the case of geometric method. 
However, key differences from pQCD calculations emerge in the details of this behavior. Referring to figure~\ref{fig:cmsatlaslhcb_a4}, the $q\bar{q}$ and $qg$ subprocesses for geometric method prediction by fitting ATLAS and LHCb data obtain reverse roles relative to pQCD predictions. Additionally, both subprocesses contribute 
with comparable weight to the $A_4$ value.  
However, comparing theoretical approaches with CMS data, we see that 
the hierarchy of the subprocess contributions 
in geometric approach aligns with predictions of pQCD. 
A key observation is that the geometric curve, when fitted to the CMS and ATLAS data within the same rapidity range $0 < |y| < 1$, 
shows significantly different ratios of the $q\bar{q}$ and $qg$ subprocess contributions. This inconsistency highlights a notable limitation of the geometric method. 
The LHCb data here presented by analysis of $\Delta A_4$ that is defined as $A_4$ minus its mean value. Notably, $\Delta A_4$ is the only dataset from the LHCb Collaboration where both pQCD and geometric method predictions demonstrate at least moderate agreement with LHC experiments. This data include a wide diapason of rapidity. Both approaches show agreement, but geometric method is sensitive to data fitting. 

In figure~\ref{fig:FB} we also show the $Q$ behavior of the $A_{\rm FB}$ asymmetry 
parameter~(\ref{Afb_Acon}) and compare our NLO predictions with statistical analysis with 
CMS data~\cite{CMS:2016bil} at $\sqrt{s}=8$ in two rapidity regions  $0<|y|<1$ and $1.25<|y|<1.5$.

\newpage

\section{Summary}
\label{Sec4_Summary}

In this paper, we analyzed and  made a comparison of the predictions for the DY angular coefficients $A_0$, $A_1$, $A_3$ and $A_4$ 
in two theoretical approaches -- pQCD at NLO based on 
the collinear factorization~\cite{Boer:2006eq,Berger:2007jw,Lyubovitskij:2024jlb}  
and geometric method~\cite{Argyres:1982kg,Chang:2017kuv,Lyu:2020nul} with 
LHC data collected by the 
CMS~\cite{CMS:2015cyj}, ATLAS~\cite{ATLAS:2016rnf}, and LHCb~\cite{LHCb:2022tbc} 
experiments across various rapidity and kinematical regions 
and employing the reduced $\chi^2$ statistical test.  
We found that the $\chi^2$ statistical test 
shows an approximate consistency of both theoretical methods.
The $Q_T$ behavior of the $A_0$ and $A_4$ coefficients is similar 
in both approaches. However, these methods disagree 
in the predictions for the angular coefficients $A_1$ and $A_3$ 
at high $Q_T$ region, even though their statistical outcomes remain comparable. These angular coefficients are related to single spin-flip hadronic structure functions.
Notably, significant limitations of the geometric method were identified, stemming from its reliance on a fitting procedure central to the approach. It can not predict the angular coefficient’s behavior in a specific rapidity range without experimental data to guide the fit. Moreover, the geometric approach can yield noticeably different ratios of quark-antiquark and quark-gluon contributions to the total angular coefficient when parameters are fitted to different experimental datasets within the same rapidity range. 

Poor agreement for the angular coefficients involving single spin-flip hadronic structure functions can be explained by the fact that the simple rotation done in geometrical method is not sensitive to rapidity and does not reproduce behavior for different contributions. In near future we plan to study full rapidity
dependence of the angular coefficients occurring in the DY process and extend our analysis to the $\alpha_s^2$ order in strong coupling expansion. 

\begin{acknowledgments}

This work was funded by ANID$-$Millen\-nium Program$-$ICN2019\_044 (Chile),  by FONDECYT (Chile) under Grant No. 1240066. 

\end{acknowledgments}

\bibliographystyle{JHEP}
\bibliography{biblio}

\providecommand{\href}[2]{#2}\begingroup\raggedright\begin{thebibliography}{10}

\bibitem{CMS:2015cyj}
{\scshape CMS} collaboration, \emph{{Angular coefficients of Z bosons produced
  in pp collisions at $\sqrt{s}$ = 8 TeV and decaying to $\mu^+ \mu^-$ as a
  function of transverse momentum and rapidity}},
  \href{https://doi.org/10.1016/j.physletb.2015.08.061}{\emph{Phys. Lett. B}
  {\bfseries 750} (2015) 154}
  [\href{https://arxiv.org/abs/1504.03512}{{\ttfamily 1504.03512}}].

\bibitem{ATLAS:2016rnf}
{\scshape ATLAS} collaboration, \emph{{Measurement of the angular coefficients
  in $Z$-boson events using electron and muon pairs from data taken at
  $\sqrt{s}=8$ TeV with the ATLAS detector}},
  \href{https://doi.org/10.1007/JHEP08(2016)159}{\emph{JHEP} {\bfseries 08}
  (2016) 159} [\href{https://arxiv.org/abs/1606.00689}{{\ttfamily
  1606.00689}}].

\bibitem{LHCb:2022tbc}
{\scshape LHCb} collaboration, \emph{{First Measurement of the $Z \rightarrow
  \mu^+ \mu^-$ Angular Coefficients in the Forward Region of pp Collisions at
  $\sqrt{s}$=13\,\,TeV}},
  \href{https://doi.org/10.1103/PhysRevLett.129.091801}{\emph{Phys. Rev. Lett.}
  {\bfseries 129} (2022) 091801}
  [\href{https://arxiv.org/abs/2203.01602}{{\ttfamily 2203.01602}}].

\bibitem{COMPASS:2017jbv}
{\scshape COMPASS} collaboration, \emph{{First measurement of
  transverse-spin-dependent azimuthal asymmetries in the Drell-Yan process}},
  \href{https://doi.org/10.1103/PhysRevLett.119.112002}{\emph{Phys. Rev. Lett.}
  {\bfseries 119} (2017) 112002}
  [\href{https://arxiv.org/abs/1704.00488}{{\ttfamily 1704.00488}}].

\bibitem{COMPASS:2023vqt}
{\scshape COMPASS} collaboration, \emph{{Final COMPASS Results on the
  Transverse-Spin-Dependent Azimuthal Asymmetries in the Pion-Induced Drell-Yan
  Process}}, \href{https://doi.org/10.1103/PhysRevLett.133.071902}{\emph{Phys.
  Rev. Lett.} {\bfseries 133} (2024) 071902}
  [\href{https://arxiv.org/abs/2312.17379}{{\ttfamily 2312.17379}}].

\bibitem{Adams:2018pwt}
B.~Adams et~al., \emph{{Letter of Intent: A New QCD facility at the M2 beam
  line of the CERN SPS (COMPASS++/AMBER)}},
  \href{https://arxiv.org/abs/1808.00848}{{\ttfamily 1808.00848}}.

\bibitem{NuSea:2006gvb}
{\scshape NuSea} collaboration, \emph{{Measurement of Angular Distributions of
  Drell-Yan Dimuons in p + d Interaction at 800-GeV/c}},
  \href{https://doi.org/10.1103/PhysRevLett.99.082301}{\emph{Phys. Rev. Lett.}
  {\bfseries 99} (2007) 082301}
  [\href{https://arxiv.org/abs/hep-ex/0609005}{{\ttfamily hep-ex/0609005}}].

\bibitem{CDF:2011ksg}
{\scshape CDF} collaboration, \emph{{First Measurement of the Angular
  Coefficients of Drell-Yan $e^{+}e^{-}$ pairs in the Z Mass Region from
  $p\bar{p}$ Collisions at $\sqrt{s}$ = 1.96 TeV}},
  \href{https://doi.org/10.1103/PhysRevLett.106.241801}{\emph{Phys. Rev. Lett.}
  {\bfseries 106} (2011) 241801}
  [\href{https://arxiv.org/abs/1103.5699}{{\ttfamily 1103.5699}}].

\bibitem{STAR:2015vmv}
{\scshape STAR} collaboration, \emph{{Measurement of the transverse single-spin
  asymmetry in $p^\uparrow+p \to W^{\pm}/Z^0$ at RHIC}},
  \href{https://doi.org/10.1103/PhysRevLett.116.132301}{\emph{Phys. Rev. Lett.}
  {\bfseries 116} (2016) 132301}
  [\href{https://arxiv.org/abs/1511.06003}{{\ttfamily 1511.06003}}].

\bibitem{PHENIX:2018dwt}
{\scshape PHENIX} collaboration, \emph{{Measurements of $\mu\mu$ pairs from
  open heavy flavor and Drell-Yan in $p+p$ collisions at $\sqrt{s}=200$ GeV}},
  \href{https://doi.org/10.1103/PhysRevD.99.072003}{\emph{Phys. Rev. D}
  {\bfseries 99} (2019) 072003}
  [\href{https://arxiv.org/abs/1805.02448}{{\ttfamily 1805.02448}}].

\bibitem{SeaQuest:2019hsx}
{\scshape SeaQuest} collaboration, \emph{{Probing nucleon\textquoteright{}s
  spin structures with polarized Drell-Yan in the Fermilab SpinQuest
  experiment}}, \href{https://doi.org/10.22323/1.346.0164}{\emph{PoS}
  {\bfseries SPIN2018} (2019) 164}
  [\href{https://arxiv.org/abs/1901.09994}{{\ttfamily 1901.09994}}].

\bibitem{Accardi:2012qut}
A.~Accardi et~al., \emph{{Electron Ion Collider: The Next QCD Frontier}:
  {Understanding the glue that binds us all}},
  \href{https://doi.org/10.1140/epja/i2016-16268-9}{\emph{Eur. Phys. J. A}
  {\bfseries 52} (2016) 268} [\href{https://arxiv.org/abs/1212.1701}{{\ttfamily
  1212.1701}}].

\bibitem{Alekhin:2024mrq}
S.~Alekhin et~al., \emph{{Status of QCD precision predictions for Drell-Yan
  processes}},  \href{https://arxiv.org/abs/2405.19714}{{\ttfamily
  2405.19714}}.

\bibitem{Boer:2006eq}
D.~Boer and W.~Vogelsang, \emph{{Drell-Yan lepton angular distribution at small
  transverse momentum}},
  \href{https://doi.org/10.1103/PhysRevD.74.014004}{\emph{Phys. Rev. D}
  {\bfseries 74} (2006) 014004}
  [\href{https://arxiv.org/abs/hep-ph/0604177}{{\ttfamily hep-ph/0604177}}].

\bibitem{Berger:2007jw}
E.L.~Berger, J.-W.~Qiu and R.A.~Rodriguez-Pedraza, \emph{{Transverse momentum
  dependence of the angular distribution of the Drell-Yan process}},
  \href{https://doi.org/10.1103/PhysRevD.76.074006}{\emph{Phys. Rev. D}
  {\bfseries 76} (2007) 074006}
  [\href{https://arxiv.org/abs/0708.0578}{{\ttfamily 0708.0578}}].

\bibitem{Lyubovitskij:2024jlb}
V.E.~Lyubovitskij, A.S.~Zhevlakov and I.A.~Anikin, \emph{{Transverse momentum
  dependence of the T-even hadronic structure functions in the Drell-Yan
  process}}, \href{https://doi.org/10.1103/PhysRevD.110.074028}{\emph{Phys.
  Rev. D} {\bfseries 110} (2024) 074028}
  [\href{https://arxiv.org/abs/2408.01243}{{\ttfamily 2408.01243}}].

\bibitem{Mirkes:1992hu}
E.~Mirkes, \emph{{Angular decay distribution of leptons from W bosons at NLO in
  hadronic collisions}},
  \href{https://doi.org/10.1016/0550-3213(92)90046-E}{\emph{Nucl. Phys. B}
  {\bfseries 387} (1992) 3}.

\bibitem{Lambertsen:2016wgj}
M.~Lambertsen and W.~Vogelsang, \emph{{Drell-Yan lepton angular distributions
  in perturbative QCD}},
  \href{https://doi.org/10.1103/PhysRevD.93.114013}{\emph{Phys. Rev. D}
  {\bfseries 93} (2016) 114013}
  [\href{https://arxiv.org/abs/1605.02625}{{\ttfamily 1605.02625}}].

\bibitem{Hamberg:1990np}
R.~Hamberg, W.L.~van Neerven and T.~Matsuura, \emph{{A complete calculation of
  the order $\alpha_s^{2}$ correction to the Drell-Yan $K$ factor}},
  \href{https://doi.org/10.1016/0550-3213(91)90064-5}{\emph{Nucl. Phys. B}
  {\bfseries 359} (1991) 343}.

\bibitem{vanNeerven:1991gh}
W.L.~van Neerven and E.B.~Zijlstra, \emph{{The $O(\alpha_s^2)$ corrected
  Drell-Yan $K$ factor in the DIS and MS scheme}},
  \href{https://doi.org/10.1016/0550-3213(92)90078-P}{\emph{Nucl. Phys. B}
  {\bfseries 382} (1992) 11}.

\bibitem{Anastasiou:2003yy}
C.~Anastasiou, L.J.~Dixon, K.~Melnikov and F.~Petriello, \emph{{Dilepton
  rapidity distribution in the Drell-Yan process at NNLO in QCD}},
  \href{https://doi.org/10.1103/PhysRevLett.91.182002}{\emph{Phys. Rev. Lett.}
  {\bfseries 91} (2003) 182002}
  [\href{https://arxiv.org/abs/hep-ph/0306192}{{\ttfamily hep-ph/0306192}}].

\bibitem{Melnikov:2006di}
K.~Melnikov and F.~Petriello, \emph{{The $W$ boson production cross section at
  the LHC through $O(\alpha^2_s)$}},
  \href{https://doi.org/10.1103/PhysRevLett.96.231803}{\emph{Phys. Rev. Lett.}
  {\bfseries 96} (2006) 231803}
  [\href{https://arxiv.org/abs/hep-ph/0603182}{{\ttfamily hep-ph/0603182}}].

\bibitem{Catani:2009sm}
S.~Catani, L.~Cieri, G.~Ferrera, D.~de~Florian and M.~Grazzini, \emph{{Vector
  boson production at hadron colliders: a fully exclusive QCD calculation at
  NNLO}}, \href{https://doi.org/10.1103/PhysRevLett.103.082001}{\emph{Phys.
  Rev. Lett.} {\bfseries 103} (2009) 082001}
  [\href{https://arxiv.org/abs/0903.2120}{{\ttfamily 0903.2120}}].

\bibitem{Gavin:2010az}
R.~Gavin, Y.~Li, F.~Petriello and S.~Quackenbush, \emph{{FEWZ 2.0: A code for
  hadronic Z production at next-to-next-to-leading order}},
  \href{https://doi.org/10.1016/j.cpc.2011.06.008}{\emph{Comput. Phys. Commun.}
  {\bfseries 182} (2011) 2388}
  [\href{https://arxiv.org/abs/1011.3540}{{\ttfamily 1011.3540}}].

\bibitem{Gavin:2012sy}
R.~Gavin, Y.~Li, F.~Petriello and S.~Quackenbush, \emph{{W Physics at the LHC
  with FEWZ 2.1}},
  \href{https://doi.org/10.1016/j.cpc.2012.09.005}{\emph{Comput. Phys. Commun.}
  {\bfseries 184} (2013) 208}
  [\href{https://arxiv.org/abs/1201.5896}{{\ttfamily 1201.5896}}].

\bibitem{Duhr:2020seh}
C.~Duhr, F.~Dulat and B.~Mistlberger, \emph{{Drell-Yan Cross Section to Third
  Order in the Strong Coupling Constant}},
  \href{https://doi.org/10.1103/PhysRevLett.125.172001}{\emph{Phys. Rev. Lett.}
  {\bfseries 125} (2020) 172001}
  [\href{https://arxiv.org/abs/2001.07717}{{\ttfamily 2001.07717}}].

\bibitem{Baglio:2022wzu}
J.~Baglio, C.~Duhr, B.~Mistlberger and R.~Szafron, \emph{{Inclusive production
  cross sections at N$^{3}$LO}},
  \href{https://doi.org/10.1007/JHEP12(2022)066}{\emph{JHEP} {\bfseries 12}
  (2022) 066} [\href{https://arxiv.org/abs/2209.06138}{{\ttfamily
  2209.06138}}].

\bibitem{Camarda:2021jsw}
S.~Camarda, L.~Cieri and G.~Ferrera, \emph{{Fiducial perturbative power
  corrections within the $\mathbf{q}_T$ subtraction formalism}},
  \href{https://doi.org/10.1140/epjc/s10052-022-10510-x}{\emph{Eur. Phys. J. C}
  {\bfseries 82} (2022) 575}
  [\href{https://arxiv.org/abs/2111.14509}{{\ttfamily 2111.14509}}].

\bibitem{Baur:2001ze}
U.~Baur, O.~Brein, W.~Hollik, C.~Schappacher and D.~Wackeroth,
  \emph{{Electroweak radiative corrections to neutral current Drell-Yan
  processes at hadron colliders}},
  \href{https://doi.org/10.1103/PhysRevD.65.033007}{\emph{Phys. Rev. D}
  {\bfseries 65} (2002) 033007}
  [\href{https://arxiv.org/abs/hep-ph/0108274}{{\ttfamily hep-ph/0108274}}].

\bibitem{Dittmaier:2001ay}
S.~Dittmaier and M.~Kr\"amer, \emph{{Electroweak radiative corrections to W
  boson production at hadron colliders}},
  \href{https://doi.org/10.1103/PhysRevD.65.073007}{\emph{Phys. Rev. D}
  {\bfseries 65} (2002) 073007}
  [\href{https://arxiv.org/abs/hep-ph/0109062}{{\ttfamily hep-ph/0109062}}].

\bibitem{Arbuzov:2005dd}
A.~Arbuzov, D.~Bardin, S.~Bondarenko, P.~Christova, L.~Kalinovskaya, G.~Nanava
  et~al., \emph{{One-loop corrections to the Drell-Yan process in SANC. I. The
  Charged current case}},
  \href{https://doi.org/10.1140/epjc/s2006-02505-y}{\emph{Eur. Phys. J. C}
  {\bfseries 46} (2006) 407}
  [\href{https://arxiv.org/abs/hep-ph/0506110}{{\ttfamily hep-ph/0506110}}].

\bibitem{Arbuzov:2007db}
A.~Arbuzov, D.~Bardin, S.~Bondarenko, P.~Christova, L.~Kalinovskaya, G.~Nanava
  et~al., \emph{{One-loop corrections to the Drell--Yan process in SANC. (II).
  The Neutral current case}},
  \href{https://doi.org/10.1140/epjc/s10052-008-0531-8}{\emph{Eur. Phys. J. C}
  {\bfseries 54} (2008) 451} [\href{https://arxiv.org/abs/0711.0625}{{\ttfamily
  0711.0625}}].

\bibitem{CarloniCalame:2007cd}
C.M.~Carloni~Calame, G.~Montagna, O.~Nicrosini and A.~Vicini, \emph{{Precision
  electroweak calculation of the production of a high transverse-momentum
  lepton pair at hadron colliders}},
  \href{https://doi.org/10.1088/1126-6708/2007/10/109}{\emph{JHEP} {\bfseries
  10} (2007) 109} [\href{https://arxiv.org/abs/0710.1722}{{\ttfamily
  0710.1722}}].

\bibitem{Gauld:2017tww}
R.~Gauld, A.~Gehrmann-De~Ridder, T.~Gehrmann, E.W.N.~Glover and A.~Huss,
  \emph{{Precise predictions for the angular coefficients in Z-boson production
  at the LHC}}, \href{https://doi.org/10.1007/JHEP11(2017)003}{\emph{JHEP}
  {\bfseries 11} (2017) 003}
  [\href{https://arxiv.org/abs/1708.00008}{{\ttfamily 1708.00008}}].

\bibitem{Martin:2004dh}
A.D.~Martin, R.G.~Roberts, W.J.~Stirling and R.S.~Thorne, \emph{{Parton
  distributions incorporating QED contributions}},
  \href{https://doi.org/10.1140/epjc/s2004-02088-7}{\emph{Eur. Phys. J. C}
  {\bfseries 39} (2005) 155}
  [\href{https://arxiv.org/abs/hep-ph/0411040}{{\ttfamily hep-ph/0411040}}].

\bibitem{Bellm:2015jjp}
J.~Bellm et~al., \emph{{Herwig 7.0/Herwig++ 3.0 release note}},
  \href{https://doi.org/10.1140/epjc/s10052-016-4018-8}{\emph{Eur. Phys. J. C}
  {\bfseries 76} (2016) 196}
  [\href{https://arxiv.org/abs/1512.01178}{{\ttfamily 1512.01178}}].

\bibitem{Li:2012wna}
Y.~Li and F.~Petriello, \emph{{Combining QCD and electroweak corrections to
  dilepton production in FEWZ}},
  \href{https://doi.org/10.1103/PhysRevD.86.094034}{\emph{Phys. Rev. D}
  {\bfseries 86} (2012) 094034}
  [\href{https://arxiv.org/abs/1208.5967}{{\ttfamily 1208.5967}}].

\bibitem{Mirkes:1994eb}
E.~Mirkes and J.~Ohnemus, \emph{{$W$ and $Z$ polarization effects in hadronic
  collisions}}, \href{https://doi.org/10.1103/PhysRevD.50.5692}{\emph{Phys.
  Rev. D} {\bfseries 50} (1994) 5692}
  [\href{https://arxiv.org/abs/hep-ph/9406381}{{\ttfamily hep-ph/9406381}}].

\bibitem{Lam:1978pu}
C.S.~Lam and W.-K.~Tung, \emph{{A Systematic Approach to Inclusive Lepton Pair
  Production in Hadronic Collisions}},
  \href{https://doi.org/10.1103/PhysRevD.18.2447}{\emph{Phys. Rev. D}
  {\bfseries 18} (1978) 2447}.

\bibitem{Accomando:2019vqt}
E.~Accomando et~al., \emph{{PDF Profiling Using the Forward-Backward Asymmetry
  in Neutral Current Drell-Yan Production}},
  \href{https://doi.org/10.1007/JHEP10(2019)176}{\emph{JHEP} {\bfseries 10}
  (2019) 176} [\href{https://arxiv.org/abs/1907.07727}{{\ttfamily
  1907.07727}}].

\bibitem{Alekhin:2014irh}
S.~Alekhin et~al., \emph{{HERAFitter}},
  \href{https://doi.org/10.1140/epjc/s10052-015-3480-z}{\emph{Eur. Phys. J. C}
  {\bfseries 75} (2015) 304} [\href{https://arxiv.org/abs/1410.4412}{{\ttfamily
  1410.4412}}].

\bibitem{Amoroso:2020fjw}
S.~Amoroso, J.~Fiaschi, F.~Giuli, A.~Glazov, F.~Hautmann and O.~Zenaiev,
  \emph{{Longitudinal Z-boson polarization and the Higgs boson production cross
  section at the Large Hadron Collider}},
  \href{https://doi.org/10.1016/j.physletb.2021.136613}{\emph{Phys. Lett. B}
  {\bfseries 821} (2021) 136613}
  [\href{https://arxiv.org/abs/2012.10298}{{\ttfamily 2012.10298}}].

\bibitem{Argyres:1982kg}
E.N.~Argyres and C.S.~Lam, \emph{{Constraints on Angular Distributions of
  Dileptons in Different Frames}},
  \href{https://doi.org/10.1103/PhysRevD.26.114}{\emph{Phys. Rev. D} {\bfseries
  26} (1982) 114}.

\bibitem{Chang:2017kuv}
W.-C.~Chang, R.E.~McClellan, J.-C.~Peng and O.~Teryaev, \emph{{Dependencies of
  lepton angular distribution coefficients on the transverse momentum and
  rapidity of $Z$ bosons produced in $pp$ collisions at the LHC}},
  \href{https://doi.org/10.1103/PhysRevD.96.054020}{\emph{Phys. Rev. D}
  {\bfseries 96} (2017) 054020}
  [\href{https://arxiv.org/abs/1708.05807}{{\ttfamily 1708.05807}}].

\bibitem{Lyu:2020nul}
Y.~Lyu, W.-C.~Chang, R.E.~Mcclellan, J.-C.~Peng and O.~Teryaev, \emph{{Lepton
  angular distribution of $W$ boson productions}},
  \href{https://doi.org/10.1103/PhysRevD.103.034011}{\emph{Phys. Rev. D}
  {\bfseries 103} (2021) 034011}
  [\href{https://arxiv.org/abs/2010.01826}{{\ttfamily 2010.01826}}].

\bibitem{Collins:1977iv}
J.C.~Collins and D.E.~Soper, \emph{{Angular Distribution of Dileptons in
  High-Energy Hadron Collisions}},
  \href{https://doi.org/10.1103/PhysRevD.16.2219}{\emph{Phys. Rev. D}
  {\bfseries 16} (1977) 2219}.

\bibitem{Lyubovitskij:2024civ}
V.E.~Lyubovitskij, W.~Vogelsang, F.~Wunder and A.S.~Zhevlakov,
  \emph{{Perturbative T-odd asymmetries in the Drell-Yan process revisited}},
  \href{https://doi.org/10.1103/PhysRevD.109.114023}{\emph{Phys. Rev. D}
  {\bfseries 109} (2024) 114023}
  [\href{https://arxiv.org/abs/2403.18741}{{\ttfamily 2403.18741}}].

\bibitem{Lam:1978zr}
C.S.~Lam and W.-K.~Tung, \emph{{Structure Function Relations at Large
  Transverse Momenta in Lepton Pair Production Processes}},
  \href{https://doi.org/10.1016/0370-2693(79)90204-1}{\emph{Phys. Lett. B}
  {\bfseries 80} (1979) 228}.

\bibitem{Collins:1984kg}
J.C.~Collins, D.E.~Soper and G.F.~Sterman, \emph{{Transverse Momentum
  Distribution in Drell-Yan Pair and W and Z Boson Production}},
  \href{https://doi.org/10.1016/0550-3213(85)90479-1}{\emph{Nucl. Phys. B}
  {\bfseries 250} (1985) 199}.

\bibitem{Peng:2015spa}
J.-C.~Peng, W.-C.~Chang, R.E.~McClellan and O.~Teryaev, \emph{{Interpretation
  of Angular Distributions of $Z$-boson Production at Colliders}},
  \href{https://doi.org/10.1016/j.physletb.2016.05.035}{\emph{Phys. Lett. B}
  {\bfseries 758} (2016) 384}
  [\href{https://arxiv.org/abs/1511.08932}{{\ttfamily 1511.08932}}].

\bibitem{Faccioli:2011pn}
P.~Faccioli, C.~Lourenco, J.~Seixas and H.K.~Wohri, \emph{{Model-independent
  constraints on the shape parameters of dilepton angular distributions}},
  \href{https://doi.org/10.1103/PhysRevD.83.056008}{\emph{Phys. Rev. D}
  {\bfseries 83} (2011) 056008}
  [\href{https://arxiv.org/abs/1102.3946}{{\ttfamily 1102.3946}}].

\bibitem{ParticleDataGroup:2024cfk}
{\scshape Particle Data Group} collaboration, \emph{{Review of particle
  physics}}, \href{https://doi.org/10.1103/PhysRevD.110.030001}{\emph{Phys.
  Rev. D} {\bfseries 110} (2024) 030001}.

\bibitem{Buckley:2014ana}
A.~Buckley, J.~Ferrando, S.~Lloyd, K.~Nordstr\"om, B.~Page, M.~R\"ufenacht
  et~al., \emph{{LHAPDF6: parton density access in the LHC precision era}},
  \href{https://doi.org/10.1140/epjc/s10052-015-3318-8}{\emph{Eur. Phys. J. C}
  {\bfseries 75} (2015) 132} [\href{https://arxiv.org/abs/1412.7420}{{\ttfamily
  1412.7420}}].

\bibitem{Yan:2022pzl}
M.~Yan, T.-J.~Hou, P.~Nadolsky and C.P.~Yuan, \emph{{CT18 global PDF fit at
  leading order in QCD}},
  \href{https://doi.org/10.1103/PhysRevD.107.116001}{\emph{Phys. Rev. D}
  {\bfseries 107} (2023) 116001}
  [\href{https://arxiv.org/abs/2205.00137}{{\ttfamily 2205.00137}}].

\bibitem{Sherstnev:2007nd}
A.~Sherstnev and R.S.~Thorne, \emph{{Parton Distributions for LO Generators}},
  \href{https://doi.org/10.1140/epjc/s10052-008-0610-x}{\emph{Eur. Phys. J. C}
  {\bfseries 55} (2008) 553} [\href{https://arxiv.org/abs/0711.2473}{{\ttfamily
  0711.2473}}].

\bibitem{CMS:2016bil}
{\scshape CMS} collaboration, \emph{{Forward\textendash{}backward asymmetry of
  Drell\textendash{}Yan lepton pairs in pp collisions at $\sqrt{s} = 8$
  $\,\mathrm{TeV}$}},
  \href{https://doi.org/10.1140/epjc/s10052-016-4156-z}{\emph{Eur. Phys. J. C}
  {\bfseries 76} (2016) 325}
  [\href{https://arxiv.org/abs/1601.04768}{{\ttfamily 1601.04768}}].

\end{thebibliography}\endgroup

\end{document}